\documentclass[10pt]{article}
\newcommand{\er}{Er$^{3+}$}
\newcommand{\ti}{Ti$^{4+}$}
\newcommand{\eri}{$^{167}$Er$^{3+}$}
\newcommand{\tio}{TiO$_2$}
\newcommand{\yso}{Y$_2$SiO$_5$}

\newcommand{\degC}{$^\text{o}$C }

\usepackage[square,numbers]{natbib}
\usepackage[margin=0.75in]{geometry}
\usepackage{authblk}
\usepackage[font=small]{caption}

\usepackage{helvet, graphicx}
\usepackage{hyperref}
\usepackage{gensymb}
\usepackage[usenames]{color}
\usepackage{amsmath,amssymb}
\usepackage{bm}
\usepackage{braket}
\usepackage{ragged2e}
\usepackage{booktabs}

\begin{document}

%\title{Narrow inhomogeneous linewidths in ion-implanted erbium in TiO$_2$}
\title{Narrow optical linewidths in erbium implanted in \tio{}}

\author[1,3]{Christopher M.~Phenicie}
\author[1,3]{Paul Stevenson}
\author[1,3]{Sacha Welinski}
\author[1]{Brendon C.~Rose}
\author[1]{Abraham T.~Asfaw}
\author[2]{Robert J.~Cava}
\author[1]{Stephen A.~Lyon}
\author[1]{Nathalie P.~de Leon}
\author[1]{Jeff D. Thompson\thanks{jdthompson@princeton.edu}}

\affil[1]{Department of Electrical Engineering, Princeton University, Princeton, NJ 08544, USA}
\affil[2]{Department of Chemistry, Princeton University, Princeton, NJ 08544, USA}
\affil[3]{Contributed equally to this work}

\date{\today}

\maketitle
\begin{abstract}
Atomic and atom-like defects in the solid-state are widely explored for quantum computers, networks and sensors. Rare earth ions are an attractive class of atomic defects that feature narrow spin and optical transitions that are isolated from the host crystal, allowing incorporation into a wide range of materials. However, the realization of long electronic spin coherence times is hampered by magnetic noise from abundant nuclear spins in the most widely studied host crystals. Here, we demonstrate that \er{} ions can be introduced via ion implantation into \tio{}, a host crystal that has not been studied extensively for rare earth ions and has a low natural abundance of nuclear spins. We observe efficient incorporation of the implanted \er into the \ti site (40\% yield), and measure narrow inhomogeneous spin and optical linewidths (20 and 460~MHz, respectively) that are comparable to  bulk-doped crystalline hosts for \er. This work demonstrates that ion implantation is a viable path to studying rare earth ions in new hosts, and is a significant step towards realizing individually addressed rare earth ions with long spin coherence times for quantum technologies.
\end{abstract}

Rare earth ion impurities in crystalline hosts are a promising platform for quantum technologies, combining narrow, stable optical transitions with isolated electronic or nuclear spins. Crystals doped with ensembles of rare earth ions (REIs) have been used to demonstrate a variety of quantum memory protocols for quantum networks~\cite{Tittel2010}, demonstrating coherence times of hours~\cite{Zhong2015}, light-matter entanglement~\cite{Clausen2011} and quantum state teleportation~\cite{Bussieres2014}. Recent work has focused on individually addressed REIs~\cite{Kolesov2012, Nakamura2014, Eichhammer2015}, and has made significant steps towards spin-photon entanglement with single ions, including the demonstration of radiatively broadened optical transitions~\cite{Zhong2018} and single-shot spin readout~\cite{Raha2019, Kindem2019} in optical nanocavities. Other efforts have focused on quantum transduction from microwave to optical frequencies ~\cite{Williamson2014, OBrien2014, Fernandez-Gonzalvo2015}. Among several widely studied REIs, erbium is particularly well suited to many of these tasks, as its telecom-wavelength optical transition enables low-loss propagation in optical fibers and integration with silicon nanophotonics~\cite{Dibos2018}.

There are several materials challenges to future development of REI-based quantum technologies. First, abundant nuclear spins in the host crystal limit REI electronic spin coherence times, despite work on several mitigation strategies including using clock states in hyperfine isotopes~\cite{McAuslan2012, Ortu2018} or electronic states with small magnetic moments~\cite{Lim2018}. The vast majority of studied hosts for REIs involve at least one element with no stable spin-0 nuclear isotopes, including all yttrium-based hosts [\emph{e.g.} {\yso} (YSO), Y$_3$Al$_5$O$_{12}$ (YAG), YLiF$_4$ and Y$_2$O$_3$], as well as alkali halides and lithium niobate (a notable exception is CaWO$_4$~\cite{Rakhmatullin2009}). Second, it is difficult to isolate single REIs in the well-studied yttrium-based materials, because they are typically contaminated with ppm-level concentrations of all rare earth elements~\cite{Liu2005,Dibos2018}. Finally, the incorporation of REIs into crystalline hosts typically involves bulk doping during growth, which precludes controllable positioning and the isolation of individual defects.

%need to cite Pr:YAG work as well
Ion implantation is an established route to controllably introducing small numbers of defects into a wide range of host materials, and in the context of quantum emitters has found recent application creating isolated color centers in diamond~\cite{Toyli2010, Chu2014, Rose2018, Sipahigil2016}. However, ion implantation has not been extensively studied in the context of REIs in crystalline hosts, particularly in the low-density regime. Ion-implanted Er$^{3+}$:YSO~\cite{Probst2014} and Gd$^{3+}$:Al$_{2}$O$_{3}$ \cite{Wisby2016} have been studied in electron spin resonance (ESR), while ion-implanted Ce$^{3+}$:YAG~\cite{Xia2015, Kornher2016} and Pr$^{3+}$:YAG~\cite{Groot-Berning2019} have been studied in single-center confocal microscopy. In these works, increased linewidths were observed compared to bulk-doped crystals, presumably as a result of unrepaired lattice damage resulting from ion implantation. In contrast, localized \er{} doping in LiNbO$_3$ using solid diffusion has yielded bulk-like properties~\cite{Thiel2012}.

In this work, we study the properties of implanted \er{} in single crystal rutile \tio{}. Based on its elemental composition, this host material is largely free of trace REI impurities and has a low abundance of nuclear spins (Ti: 87\% I=0, O: 99.96\% I=0), which could be reduced further using isotopically enriched Ti precursors. Bulk-doped \er{}:\tio{} has previously been observed in electron spin resonance (ESR) in the context of maser development~\cite{Gerritsen1962, Sabisky1966}. Here, we use ESR and photoluminescence (PL) spectroscopy to demonstrate that implanted \er{} ions in \tio{} have inhomogeneous spin and optical linewidths (20~MHz and 460~MHz, respectively) that are comparable to or smaller than typical values for bulk-doped \er{} in most host crystals~\cite{Thiel2011}. The \er{} ions that we observe occupy substitutional Ti$^{4+}$ sites with D$_{2h}$ symmetry, and we hypothesize that the narrow linewidths are a consequence of the non-polar symmetry of this site. We determine the energy of the lowest few ground and excited state crystal field levels. For the sample processing conditions that lead to the highest yield, we find that $40 \pm 16$~\% of the implanted \er{} ions are situated in \ti{} sites.

Samples of rutile \tio{} (MTI) with (001) orientation were implanted with erbium ions with a range of energies up to 350~keV chosen to achieve a uniform \er{} concentration in the first 100~nm of the crystal, and total fluences of 9$\times\ 10^{11}$ cm$^{-2}$ (hereafter, ``low dose'') and 9$\times\ 10^{13}$ cm$^{-2}$ (high dose). After implantation, some samples were annealed in air at temperatures up to 1000\degC{} (Table \ref{tab:samples} and Ref.~\cite{suppinfo}). We look for evidence of implanted erbium using photoluminescence excitation (PLE) spectroscopy in a helium cryostat, performed by scanning a chopped laser and collecting delayed fluorescence onto a low-noise photodiode~\cite{Dibos2018}. The resulting spectrum (Figure 1a) shows several sharp absorption resonances near 1.5 $\mu$m, with the narrowest having a total linewidth of 460~MHz (Figure 1b). During the scan, we measure the fluorescence lifetime at each excitation wavelength (Figure 1c) and find that the four starred peaks have the same lifetime ($5.25 \pm 0.03$~ms), while some of the smaller peaks have a shorter lifetime ($2.51 \pm 0.04$~ms).

\begin{figure}[t!]
  \centering
  \includegraphics[width = 3.25in]{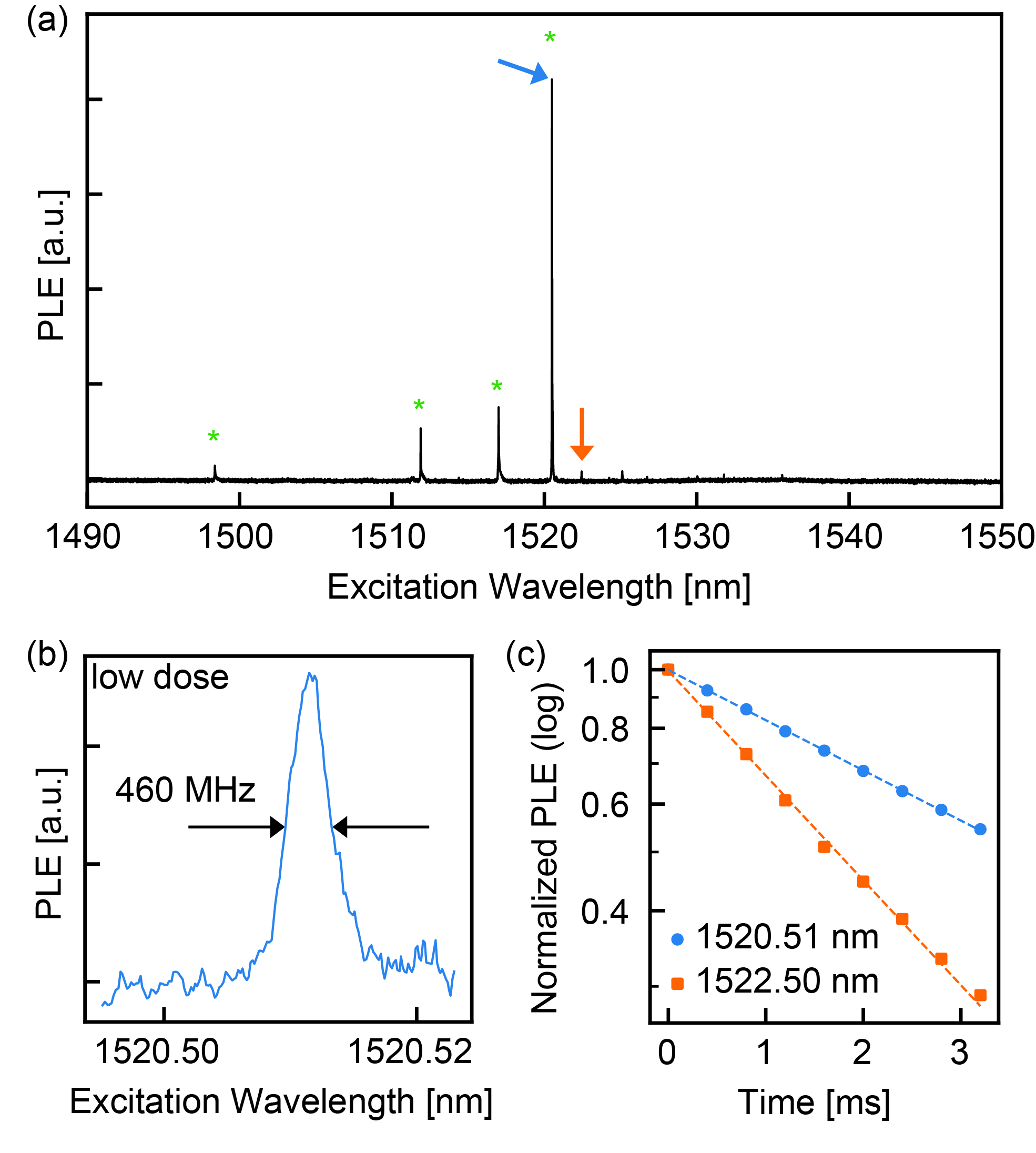}
  \caption{
  %Optical absorption and fluorescence lifetime of \er{}:\tio{}. 
  (a) PLE spectrum at 11~K in a high-dose sample. Stars indicate lines belonging to $Z_1 \rightarrow Y_1 - Y_4$ transitions of the same \er site, as determined by PL (see Figure 2).
  (b) Spectrum at 5~K of the $Z_1-Y_1$ transition in a low-dose sample, demonstrating the narrowest observed inhomogeneous linewidth of 460~MHz (FWHM).
  (c) Fluorescence lifetime measurements under excitation at the wavelengths indicated by arrows in (a). The long (blue dots) lifetime is 5.25 $\pm$ 0.03~ms, while the short (orange squares) lifetime is 2.51 $\pm$ 0.04~ms. The other starred peaks in (a) exhibit the same lifetime as the blue curve.
  }
  
  \label{fig:Fig1}
\end{figure}

In the site symmetries possible in the rutile space group $P4_2/mnm$, the ground ($^4I_{15/2}$) and excited ($^4I_{13/2}$) electronic states of \er{} split into 8 and 7 Kramers' doublets labeled $Z_{1-8}$ and $Y_{1-7}$, respectively (Figure 2a). Absorption lines arise from transitions from $Z_1 \rightarrow Y_n$, while fluorescence occurs primarily from $Y_1 \rightarrow Z_n$, regardless of which $Y$ level is excited, because of rapid nonradiative relaxation to $Y_1$ \cite{Liu2005, Bottger2006}. We conjecture that the four starred peaks in Figure 1a correspond to transitions from $Z_1 \rightarrow Y_1-Y_4$ in the same \er{} site. To check this hypothesis, we filter the fluorescence through a grating spectrometer to resolve the emission wavelength in a PL measurement (Figure 2b) while exciting at each of these peaks. The emission spectra are qualitatively similar, with a principal peak at 1520~nm and a series of smaller peaks at longer wavelengths, confirming this hypothesis. Furthermore, the absence of any shorter-wavelength emission confirms that 1520~nm is the $Z_1 \rightarrow Y_1$ transition.

The emission spectrum also allows several of the $Z$ energy levels to be determined. These measurements are performed at $T= 11$~K to increase the excitation efficiency by homogeneously broadening the optical transition, but at these temperatures, several excited $Y$ states are appreciably thermally populated and contribute extra lines to the emission spectrum. These transitions can be separated using temperature-dependent PL spectroscopy (Figure 2c), where lines with the same temperature dependence are interpreted to originate from the same $Y_i$ level (Figure 2d). This allows us to assign energies to the first five ground states $Z_1 - Z_5$, and confirms the first three $Y$ assignments determined from Figure 1a. We additionally note that the PL line intensities show that the dominant decay pathways for the first few excited states are $Y_1$ to $Z_1$, $Y_2$ to $Z_2$, and $Y_3$ to $Z_3$ and $Z_4$.

\begin{figure}[t!]
  \centering
  \includegraphics[width = 3.25in]{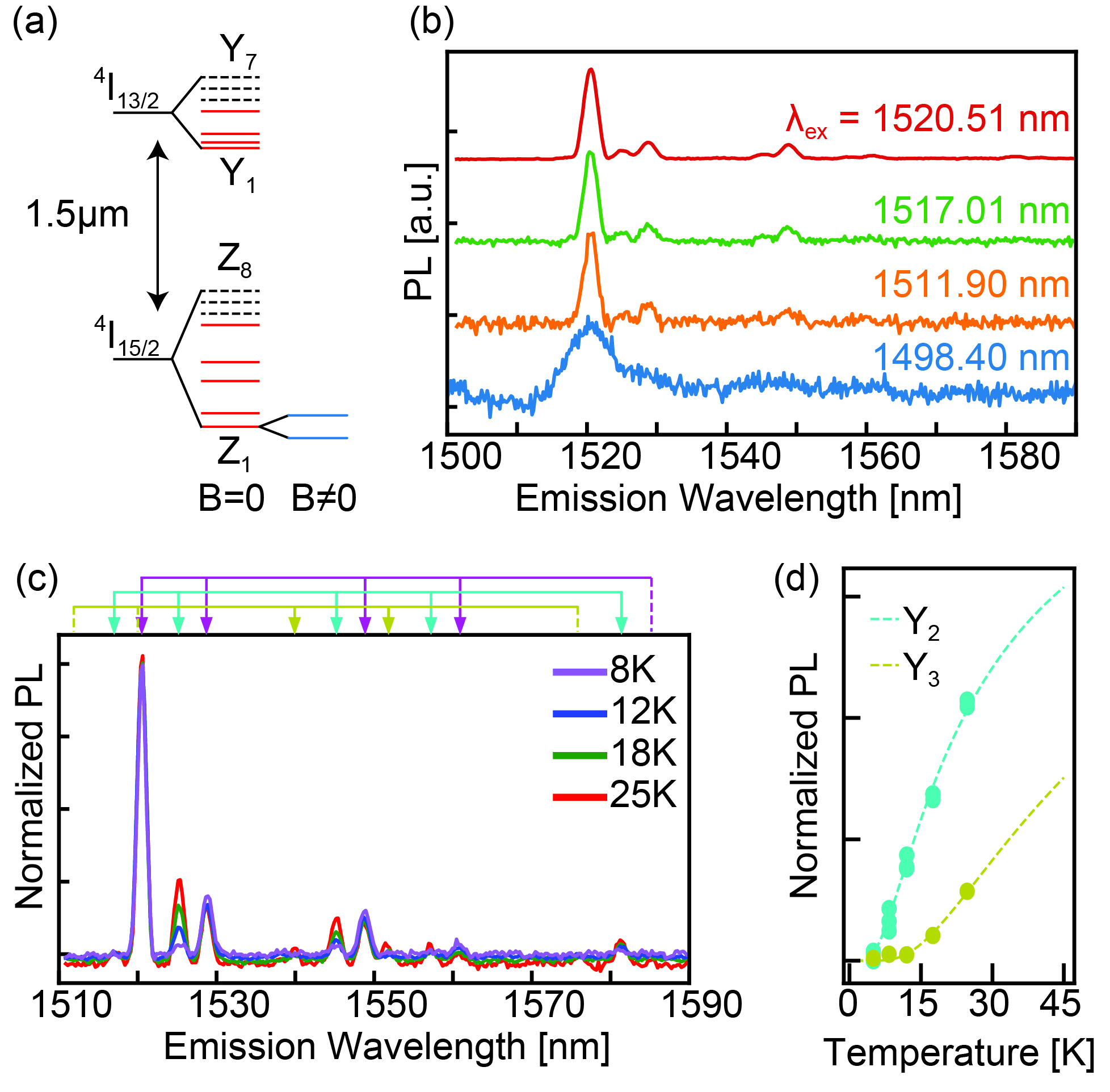}
  \caption{
  %Emission spectrum of \er{}:\tio{}
  (a) Level structure of \er{}:\tio{}. Solid red lines indicate crystal field levels in the $^4I_{15/2}$ and $^4I_{13/2}$ manifolds measured in this work (also listed in Table \ref{tab:energyLevels}). Unobserved levels are depicted with dashed lines. Each line is a Kramers' doublet, which splits into two sublevels when a magnetic field is applied.
  (b) Emission spectrum observed when exciting on the $Z_1 \rightarrow Y_1 - Y_4$ transitions at the indicated wavelength $\lambda_{\text{ex}}$. The linewidths are  instrument-limited, with a wider slit width for the 1498.40~nm measurement, since the fluorescence was very weak.
  (c) Emission spectrum with $Z_1 \rightarrow Y_1$ excitation at several temperatures, normalized to the $Y_1 \rightarrow Z_1$ peak. The magenta, cyan, and yellow arrows indicate groupings of lines corresponding to decay to $Z_1 - Z_5$ from $Y_1, Y_2$ and $Y_3$, respectively. Lines without arrowheads are not observed.
  (d) Temperature dependence of the PL intensity for different lines in (c). Dashed lines (cyan, yellow, corresponding to arrows of the same color in (c)) are the expected PL intensities as a function of temperature assuming thermal equilibrium within the $Y$ manifold.
  }
  
  \label{fig:Fig2}
\end{figure}

\begin{table}
	\centering
    % \justify
	\setlength\tabcolsep{0.14cm}
	\begin{tabular}{rcrrr}
        \toprule
        Term & Level & & Energy & \\
          &  &      THz &       nm (vac.) & cm$^{-1}$ \\
        \midrule
         $^4I_{15/2}$ &    $Z_1$ &        0 &        - &         0 \\
                      &    $Z_2$ &     1.06 &        - &      35.3 \\
                      &    $Z_3$ &     3.61 &        - &       120 \\
                      &    $Z_4$ &     5.10 &        - &       170 \\
                      &    $Z_5$ &     8.05 &        - &       268 \\
                      &    $Z_6$-$Z_8$ &         &       not observed &          \\
%                      &    $Z_7$ &        - &        - &         - \\
%                      &    $Z_8$ &        - &        - &         - \\
        \\
         $^4I_{13/2}$ &    $Y_1$ & 197.1651 &  1520.515 &   6576.719 \\
                      &    $Y_2$ & 197.6204 &  1517.012 &   6591.905 \\
                      &    $Y_3$ & 198.2888 &  1511.898 &   6614.203 \\
                      &    $Y_4$ & 200.0745 &  1498.404 &   6673.767 \\
                      &    $Y_5$-$Y_7$ &         &        not observed &          \\
%                      &    $Y_6$ &        - &        - &         - \\
%                      &    $Y_7$ &        - &        - &         - \\
        \bottomrule
    \end{tabular}
    \caption{Energy levels for \er{}:TiO$_2$. The uncertainty in the $Y$ energies is 200~MHz, determined from a wavemeter calibrated to an acetylene absorption cell. The uncertainty in the $Z$ levels is $\sim$ 10~GHz, based on fits to the data in Figure 2b.}

	\label{tab:energyLevels}
\end{table}

Next, we probe the spin properties of the $Z_1$ doublet using ESR in an X-band continuous wave (CW) spectrometer. We identify the \er{} peak through its characteristic high $g$-factor ($g_{zz}=14.30 \pm 0.42$) and hyperfine spectrum (\eri{} with $I=7/2$ and 23\% abundance, $A_{zz} = 1503 \pm 11$~MHz, Figure 3a). The magnetic moment varies strongly with the magnetic field orientation, with its maximal value along the $[001]$ direction ($c$-axis). The effective g-factor in the $[100]$ direction is $1.63 \pm 0.3$, as determined from a fit to the orientation dependence (Figure 3b). We are unable to directly measure the line in this orientation because it is very weak. We note that these $g$-factors are not consistent with those previously reported for \er:\tio\ ($g_{zz} = 15.1$, $g_{xx},g_{yy} < 0.1$)~\cite{Gerritsen1962}, although the hyperfine constant is similar to the previously reported value ($A = 1484$~MHz).

To probe whether the \er{} site identified in ESR is the same as the one in the optical measurements above, we illuminate the sample with light near the $Z_1 \rightarrow Y_1$ line and observe the resulting change in the ESR signal. The experimental apparatus for this measurement is depicted in Figure 3c. The ESR signal changes when the laser is resonant with the $Z_1 \rightarrow Y_1$ optical transition (Figure 3d) because some fraction of the \er{} ions are transferred to the $Y_1$ state and are no longer resonant with the microwaves. This confirms that the optical and ESR measurements probe the same \er{} site. In a magnetic field, the $Z_1\rightarrow Y_1$ transition splits into four lines. Here, only two are observed, presumably because the other two are weak or orthogonal to the light polarization direction. From the splitting, we can extract the $Y_1$ magnetic moment along the $c$-axis, $g^e_{zz}$: the spacing between the peaks corresponds to a difference in $g$-factors $|g^g_{zz} - g^e_{zz}| = 2.105 \pm 0.013$, implying $g^e_{zz} = 12.19 \pm 0.42 < g^g_{zz}$, since the larger value is bigger than the maximum possible value of $14.4$ for $^4I_{13/2}$.

\begin{figure}[h!]
  \centering
  \includegraphics[width = 3.25in]{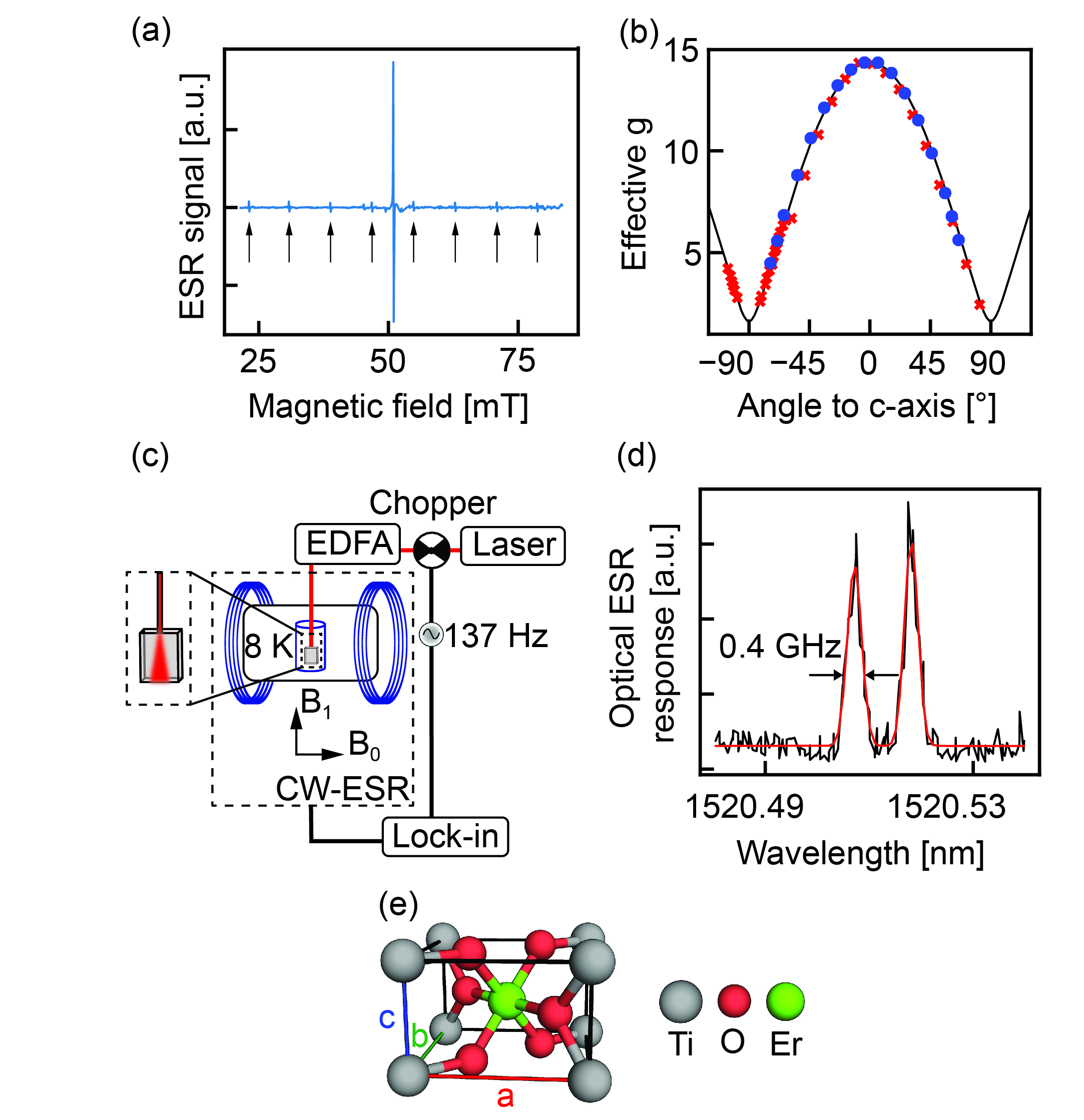}
  \caption{
  %Spin properties of \er{}:\tio{} and correlations with the optical spectrum. 
  (a) CW ESR spectrum with the magnetic field nearly parallel to the $c$-axis of the crystal. Arrows point to hyperfine lines in the spectrum belonging to the \eri{} ions.
  (b) Angular variation of the effective $g$-factor of \er{} in the $ac$- (red crosses) and $bc$- (blue circles) planes.
  (c) Schematic of the optical-ESR setup. The sample sits in a CW ESR spectrometer, and is illuminated by a laser delivered through an optical fiber. When the laser is resonant with the $Z_1 \rightarrow Y_1$ optical transition, we observe a change in the ESR signal from population shelving in the excited state. To increase sensitivity, we measure this optical ESR response by chopping the excitation light and measuring the modulated ESR signal via lock-in detection.
  (d) The optical ESR response is resonant at the same wavelength as the $Z_1$-$Y_1$ spectral line in Figure 1, indicating that the spin and optical transitions arise from the same \er\ site.  
  (e) Unit cell of \tio{} (rutile). We propose that \er{} occupies a Ti site, as shown in green.
  }
  \label{fig:Fig3}
\end{figure}

The rotation dependence of $g$ is essentially identical in the $ac$ and $bc$ crystallographic planes; however, we observe that the single line splits into two lines when the field is rotated slightly into the $ab$ plane. These observations suggest that \er{} is incorporated into a well-defined crystallographic site with two orientations differing by a 90 degree rotation around the $c$-axis, which is consistent with substitutional incorporation on the \ti{} site with $D_{2h}$ symmetry. This incorporation site has been suggested using similar measurement techniques for a range of transition metal impurities in \tio{}, including iron~\cite{Carter1960}, manganese~\cite{Gerritsen1963} and others~\cite{Gerritsen1962}; in contrast, nickel incorporates in an interstitial site and gives a qualitatively different ESR spectrum~\cite{Gerritsen1962Ni}.

Lastly, we study the dependence of the ion properties on implantation and annealing conditions. First, the implantation yield into the \ti{} site (measured using quantitative ESR~\cite{suppinfo}) depends strongly on the implantation dose (Figure 4a). Without post-implantation annealing, the yield for low-dose implantation is 40 $\pm\ 16$~\%, while for high-dose implantation it is only 2 $\pm\ 1$~\%. After annealing in air for 2 h at 1000$^\text{o}$C, the yield for the high-dose sample increases to 11 $\pm\ 4$~\%, but is unchanged in the low-dose sample. The optical lineshapes also change with implantation and annealing conditions. Figure 4b shows the PLE spectrum of the $Z_1 \rightarrow Y_1$ transition in the high-dose sample as a function of annealing temperature, demonstrating a significant reduction in the overall peak width and confirming the change in yield measured with ESR. Figure 4c shows the PLE spectrum of the  $Z_1 \rightarrow Y_1$ transition in the low-dose and high-dose samples after 1000\degC{} annealing, demonstrating that the inhomogeneous linewidth is much smaller in the low-dose sample (FWHM of central feature is 0.5 vs 1.6~GHz), and that the broad, long-wavelength tail observed in all of the PLE features in the high-dose sample (Figure 4b) is absent in the low-dose sample. We note that the linewidth observed in the optical-ESR spectrum in the high-dose sample (Figure 3d) does not show a long-wavelength tail. This may result from probing only a subset of un-disturbed ions selected by the ESR transition.

\begin{figure}[h!]
  \centering
  \includegraphics[width = 3.25in]{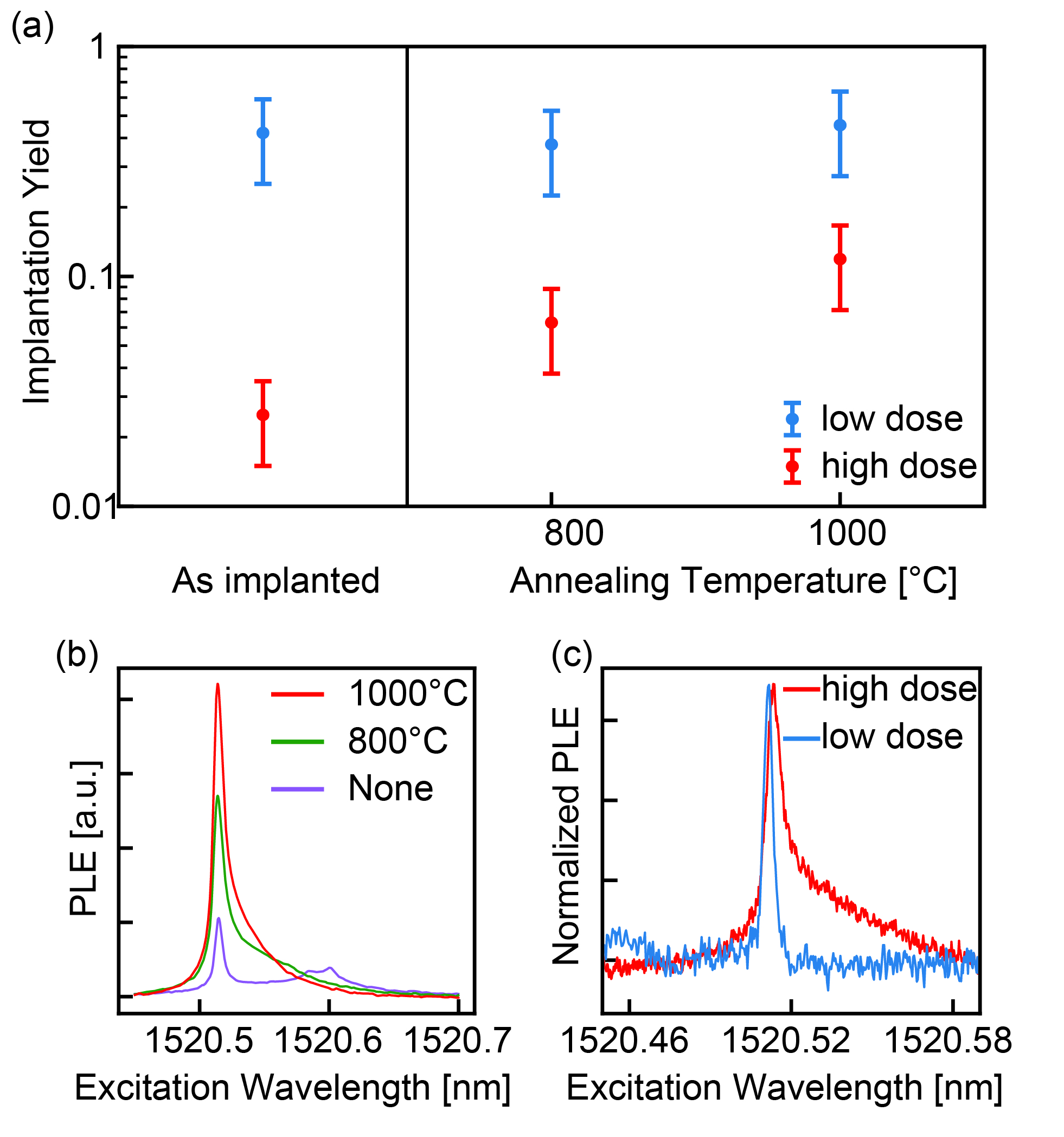}
  \caption{
  %Effects of annealing on the optical spectrum. 
  (a) Implantation yield, measured using quantitative ESR, as a function of implantation and annealing conditions.
  (b) PLE spectrum at $T = 11$~K of the $Z_1 \rightarrow Y_1$ transition in the high-dose sample after different annealing conditions.
  (c) PLE spectrum at $T = 5$~K of the $Z_1 \rightarrow Y_1$ transition in the low-dose and high-dose samples after a 1000$^\circ$C anneal. The data are scaled to have the same peak height to facilitate comparison of the linewidths.
  }
  
  \label{fig:Fig4}
\end{figure}

\begin{table*}
	\centering
% 	\begin{tabular}{|l|l|l|l|l|}
	\begin{tabular}{lllll}
	\toprule 
	\# & Total fluence                       & Annealing conditions                & Implantation yield & Data shown in Figure\\
	\hline
	1  &  $9 \times 10^{13}$ cm$^{-2}$     & As implanted                        & 2 $\pm\ 1$~\%  & \ref{fig:Fig3}a),b),\ref{fig:Fig4}b)\\ 
	2  &  "                                 & 800$^{\text{o}}$C, 2 hours in air   & 5 $\pm\ 2$~\%  &  \ref{fig:Fig4}b)\\
	3  &  "                                  & 1000$^{\text{o}}$C, 2 hours in air  & 11 $\pm\ 4$~\%  &  \ref{fig:Fig1}a), \ref{fig:Fig2}b),c),\ref{fig:Fig3}c),\ref{fig:Fig4}b),d)\\
	4  &  $9 \times 10^{11}$ cm$^{-2}$     & As implanted                        & 40 $\pm\ 16$~\% &  \\ 
	5  &  "                                & 800$^{\text{o}}$C, 2 hours in air   & 34 $\pm\ 14$~\%  &  \\
	6  &  "                                  & 1000$^{\text{o}}$C, 2 hours in air  & 40 $\pm\ 16$~\%  & \ref{fig:Fig1}b),\ref{fig:Fig4}c) \\
	\bottomrule
	\end{tabular}

    \caption{List of samples used in this work.}
	\label{tab:samples}
\end{table*}

We now turn to a discussion of these results. Several important properties follow from the symmetry of the $D_{2h}$ incorporation site. First, the optical transition should be nearly purely magnetic dipole (MD) in nature, since the inversion-symmetric crystal field cannot mix 4$f$ and 5$d$ states to introduce a forced electric dipole transition. The observed excited state lifetime of 5.25~ms is consistent with the predicted MD decay rate of 5-6.7~ms~\cite{Dodson2012}, depending on the  \tio{} refractive index for the polarization direction of the dipole ($n_o = 2.45$, $n_e = 2.70$)~\cite{Rams1997}. Second, an expected consequence of the non-polar nature of this site is that the wavefunctions should not have a permanent electric dipole moment, eliminating first-order sensitivity of the optical transitions to electric fields. We hypothesize that the observed narrow inhomogeneous linewidths are partially a consequence of this insensitivity, and that the asymmetric inhomogeneous linewidths observed in the high-dose sample (Figure 4b,c) arise from a quadratic DC Stark shift. In future work, it will be interesting to probe whether the lack of a permanent dipole moment leads to long coherence of the optical transitions.

Previous studies of ion implantation in \tio{} using Rutherford backscattering have reported nearly unity substitutional fractions of Sn and Hf implanted in \tio{} at fluences up to $3 \times 10^{15}$ cm$^{-2}$ without significant lattice damage, and that significant lattice damage from implanting La (which does not substitute \ti{}) can be reversed by annealing above 900$^\text{o}$C~\cite{Fromknecht1996}. The \er{} properties that we measure under different implantation and annealing conditions are consistent with a picture that lattice damage caused by ion implantation adversely affects the defect properties, and the amount of damage is proportional to the implantation dose and can be reduced by post-implantation thermal annealing. The dependence of the implantation yield on dose is similar to reported values for Ce$^{3+}$ and Pr$^{3+}$ in YAG~\cite{Kornher2016,Groot-Berning2019}.

An important question about the incorporation of trivalent \er{} into the \ti{} site is how charge compensation is achieved. In many materials, including alkali halides and CaF$_2$, local charge compensation in nearest or next-nearest neighbor sites around aliovalent defects is observed through a lowering of the symmetry and increase in the number of spectral lines~\cite{Watkins1959}, while in other materials, such as CaWO$_4$~\cite{Mims1968}, SrTiO$_3$~\cite{Cockroft1992} and \tio{}~\cite{Gerritsen1962}, this effect is weak or absent, and the charge compensation is believed to be long-range. Since our ESR spectra are consistent with a single crystallographic site with two orientations, we conclude that the charge compensation is remote for \er{}:\tio{}. Based on a broad ESR scan, we believe our sample has non-negligible concentrations of several transition metal impurities, and changes in their valence state may also play a role. We cannot rule out that a small minority of \er sites have a local charge-compensating defect.

We conclude with a discussion of prospects for building quantum systems out of individually addressed \er{} ions in \tio{} using nanophotonic circuits~\cite{Dibos2018, Raha2019}. The apparent strong branching ratio of $Y_1$ to $Z_1$ is favorable for Purcell enhancement of the emission, and we note that MD Purcell enhancement can be of a similar magnitude to the more common electric dipole enhancement~\cite{Dibos2018}. Furthermore, the low implantation dose used here is already quite high from the perspective of single ion studies, corresponding to an areal density of $10^4\,\mu$m$^{-2}$ and an average defect spacing of 21~nm, suggesting that high implantation yield and narrow linewidths can be expected in the relevant density range for single-ion work. The significant reduction in the concentration of nuclear magnetic moments compared to typical \er{} hosts like YSO and YAG may enable extended electronic spin coherence times and open the door to manipulating individual nuclear spins~\cite{Jiang2009}, and further reduction of the nuclear magnetism may be accomplished by CVD growth~\cite{Ghoshtagore1970} of isotopically enriched layers.

We note that these results, together with recent work on implanted color centers in diamond~\cite{Sipahigil2016, Rose2018, Chu2014},  suggest that good defect properties can be achieved using ion implantation and thermal annealing. In contrast to doping during growth, ion implantation allows rapid exploration of many materials and defects. This approach may be extended to search for other hosts for rare earth ions, as well as other optically active defects, which may be attractive for a wide range of quantum technologies including computing, communications and sensing.

\section*{Acknowledgements}
% \begin{acknowledgements}
We gratefully acknowledge helpful conversations with Philippe Goldner, early technical contributions to the apparatus from Henry Ando and Zheru Qiu, and assistance with data collection from Sabrina Chern. Funding for this research was provided by the AFOSR (contract FA9550-18-1-0334), the Eric and Wendy Schmidt Transformative Technology Fund, the Princeton Catalysis Initiative and the Princeton Center for Complex Materials (PCCM), an NSF-funded MRSEC (DMR-1420541). We acknowledge the use of Princeton's Imaging and Analysis Center, which is partially supported by PCCM, as well as the Princeton Micro-Nano Fabrication Lab. C.M.P is supported by an NDSEG graduate fellowship.
% \end{acknowledgements}

%\input{suppinfo}

\bibliography{library_er_tio2}

%merlin.mbs apsrev4-1.bst 2010-07-25 4.21a (PWD, AO, DPC) hacked
%Control: key (0)
%Control: author (72) initials jnrlst
%Control: editor formatted (1) identically to author
%Control: production of article title (-1) disabled
%Control: page (0) single
%Control: year (1) truncated
%Control: production of eprint (0) enabled
\begin{thebibliography}{45}%
\makeatletter
\providecommand \@ifxundefined [1]{%
 \@ifx{#1\undefined}
}%
\providecommand \@ifnum [1]{%
 \ifnum #1\expandafter \@firstoftwo
 \else \expandafter \@secondoftwo
 \fi
}%
\providecommand \@ifx [1]{%
 \ifx #1\expandafter \@firstoftwo
 \else \expandafter \@secondoftwo
 \fi
}%
\providecommand \natexlab [1]{#1}%
\providecommand \enquote  [1]{``#1''}%
\providecommand \bibnamefont  [1]{#1}%
\providecommand \bibfnamefont [1]{#1}%
\providecommand \citenamefont [1]{#1}%
\providecommand \href@noop [0]{\@secondoftwo}%
\providecommand \href [0]{\begingroup \@sanitize@url \@href}%
\providecommand \@href[1]{\@@startlink{#1}\@@href}%
\providecommand \@@href[1]{\endgroup#1\@@endlink}%
\providecommand \@sanitize@url [0]{\catcode `\\12\catcode `\$12\catcode
  `\&12\catcode `\#12\catcode `\^12\catcode `\_12\catcode `\%12\relax}%
\providecommand \@@startlink[1]{}%
\providecommand \@@endlink[0]{}%
\providecommand \url  [0]{\begingroup\@sanitize@url \@url }%
\providecommand \@url [1]{\endgroup\@href {#1}{\urlprefix }}%
\providecommand \urlprefix  [0]{URL }%
\providecommand \Eprint [0]{\href }%
\providecommand \doibase [0]{http://dx.doi.org/}%
\providecommand \selectlanguage [0]{\@gobble}%
\providecommand \bibinfo  [0]{\@secondoftwo}%
\providecommand \bibfield  [0]{\@secondoftwo}%
\providecommand \translation [1]{[#1]}%
\providecommand \BibitemOpen [0]{}%
\providecommand \bibitemStop [0]{}%
\providecommand \bibitemNoStop [0]{.\EOS\space}%
\providecommand \EOS [0]{\spacefactor3000\relax}%
\providecommand \BibitemShut  [1]{\csname bibitem#1\endcsname}%
\let\auto@bib@innerbib\@empty
%</preamble>
\bibitem [{\citenamefont {Tittel}\ \emph {et~al.}(2009)\citenamefont {Tittel},
  \citenamefont {Afzelius}, \citenamefont {Chaneli{\'{e}}re}, \citenamefont
  {Cone}, \citenamefont {Kr{\"{o}}ll}, \citenamefont {Moiseev},\ and\
  \citenamefont {Sellars}}]{Tittel2010}%
  \BibitemOpen
  \bibfield  {author} {\bibinfo {author} {\bibfnamefont {W.}~\bibnamefont
  {Tittel}}, \bibinfo {author} {\bibfnamefont {M.}~\bibnamefont {Afzelius}},
  \bibinfo {author} {\bibfnamefont {T.}~\bibnamefont {Chaneli{\'{e}}re}},
  \bibinfo {author} {\bibfnamefont {R.}~\bibnamefont {Cone}}, \bibinfo {author}
  {\bibfnamefont {S.}~\bibnamefont {Kr{\"{o}}ll}}, \bibinfo {author}
  {\bibfnamefont {S.}~\bibnamefont {Moiseev}}, \ and\ \bibinfo {author}
  {\bibfnamefont {M.}~\bibnamefont {Sellars}},\ }\href {\doibase
  10.1002/lpor.200810056} {\bibfield  {journal} {\bibinfo  {journal} {Laser
  {\&} Photonics Reviews}\ }\textbf {\bibinfo {volume} {4}},\ \bibinfo {pages}
  {244} (\bibinfo {year} {2009})}\BibitemShut {NoStop}%
\bibitem [{\citenamefont {Zhong}\ \emph {et~al.}(2015)\citenamefont {Zhong},
  \citenamefont {Hedges}, \citenamefont {Ahlefeldt}, \citenamefont
  {Bartholomew}, \citenamefont {Beavan}, \citenamefont {Wittig}, \citenamefont
  {Longdell},\ and\ \citenamefont {Sellars}}]{Zhong2015}%
  \BibitemOpen
  \bibfield  {author} {\bibinfo {author} {\bibfnamefont {M.}~\bibnamefont
  {Zhong}}, \bibinfo {author} {\bibfnamefont {M.~P.}\ \bibnamefont {Hedges}},
  \bibinfo {author} {\bibfnamefont {R.~L.}\ \bibnamefont {Ahlefeldt}}, \bibinfo
  {author} {\bibfnamefont {J.~G.}\ \bibnamefont {Bartholomew}}, \bibinfo
  {author} {\bibfnamefont {S.~E.}\ \bibnamefont {Beavan}}, \bibinfo {author}
  {\bibfnamefont {S.~M.}\ \bibnamefont {Wittig}}, \bibinfo {author}
  {\bibfnamefont {J.~J.}\ \bibnamefont {Longdell}}, \ and\ \bibinfo {author}
  {\bibfnamefont {M.~J.}\ \bibnamefont {Sellars}},\ }\href {\doibase
  10.1038/nature14025} {\bibfield  {journal} {\bibinfo  {journal} {Nature}\
  }\textbf {\bibinfo {volume} {517}},\ \bibinfo {pages} {177} (\bibinfo {year}
  {2015})}\BibitemShut {NoStop}%
\bibitem [{\citenamefont {Clausen}\ \emph {et~al.}(2011)\citenamefont
  {Clausen}, \citenamefont {Usmani}, \citenamefont {Bussi{\'{e}}res},
  \citenamefont {Sangouard}, \citenamefont {Afzelius}, \citenamefont {{De
  Riedmatten}},\ and\ \citenamefont {Gisin}}]{Clausen2011}%
  \BibitemOpen
  \bibfield  {author} {\bibinfo {author} {\bibfnamefont {C.}~\bibnamefont
  {Clausen}}, \bibinfo {author} {\bibfnamefont {I.}~\bibnamefont {Usmani}},
  \bibinfo {author} {\bibfnamefont {F.}~\bibnamefont {Bussi{\'{e}}res}},
  \bibinfo {author} {\bibfnamefont {N.}~\bibnamefont {Sangouard}}, \bibinfo
  {author} {\bibfnamefont {M.}~\bibnamefont {Afzelius}}, \bibinfo {author}
  {\bibfnamefont {H.}~\bibnamefont {{De Riedmatten}}}, \ and\ \bibinfo {author}
  {\bibfnamefont {N.}~\bibnamefont {Gisin}},\ }\href {\doibase
  10.1038/nature09662} {\bibfield  {journal} {\bibinfo  {journal} {Nature}\
  }\textbf {\bibinfo {volume} {469}},\ \bibinfo {pages} {508} (\bibinfo {year}
  {2011})}\BibitemShut {NoStop}%
\bibitem [{\citenamefont {Bussi{\`{e}}res}\ \emph {et~al.}(2014)\citenamefont
  {Bussi{\`{e}}res}, \citenamefont {Clausen}, \citenamefont {Tiranov},
  \citenamefont {Korzh}, \citenamefont {Verma}, \citenamefont {Nam},
  \citenamefont {Marsili}, \citenamefont {Ferrier}, \citenamefont {Goldner},
  \citenamefont {Herrmann}, \citenamefont {Silberhorn}, \citenamefont {Sohler},
  \citenamefont {Afzelius},\ and\ \citenamefont {Gisin}}]{Bussieres2014}%
  \BibitemOpen
  \bibfield  {author} {\bibinfo {author} {\bibfnamefont {F.}~\bibnamefont
  {Bussi{\`{e}}res}}, \bibinfo {author} {\bibfnamefont {C.}~\bibnamefont
  {Clausen}}, \bibinfo {author} {\bibfnamefont {A.}~\bibnamefont {Tiranov}},
  \bibinfo {author} {\bibfnamefont {B.}~\bibnamefont {Korzh}}, \bibinfo
  {author} {\bibfnamefont {V.~B.}\ \bibnamefont {Verma}}, \bibinfo {author}
  {\bibfnamefont {S.~W.}\ \bibnamefont {Nam}}, \bibinfo {author} {\bibfnamefont
  {F.}~\bibnamefont {Marsili}}, \bibinfo {author} {\bibfnamefont
  {A.}~\bibnamefont {Ferrier}}, \bibinfo {author} {\bibfnamefont
  {P.}~\bibnamefont {Goldner}}, \bibinfo {author} {\bibfnamefont
  {H.}~\bibnamefont {Herrmann}}, \bibinfo {author} {\bibfnamefont
  {C.}~\bibnamefont {Silberhorn}}, \bibinfo {author} {\bibfnamefont
  {W.}~\bibnamefont {Sohler}}, \bibinfo {author} {\bibfnamefont
  {M.}~\bibnamefont {Afzelius}}, \ and\ \bibinfo {author} {\bibfnamefont
  {N.}~\bibnamefont {Gisin}},\ }\href {\doibase 10.1038/nphoton.2014.215}
  {\bibfield  {journal} {\bibinfo  {journal} {Nature Photonics}\ }\textbf
  {\bibinfo {volume} {8}},\ \bibinfo {pages} {775} (\bibinfo {year}
  {2014})}\BibitemShut {NoStop}%
\bibitem [{\citenamefont {Kolesov}\ \emph {et~al.}(2012)\citenamefont
  {Kolesov}, \citenamefont {Xia}, \citenamefont {Reuter}, \citenamefont
  {St{\"{o}}hr}, \citenamefont {Zappe}, \citenamefont {Meijer}, \citenamefont
  {Hemmer},\ and\ \citenamefont {Wrachtrup}}]{Kolesov2012}%
  \BibitemOpen
  \bibfield  {author} {\bibinfo {author} {\bibfnamefont {R.}~\bibnamefont
  {Kolesov}}, \bibinfo {author} {\bibfnamefont {K.}~\bibnamefont {Xia}},
  \bibinfo {author} {\bibfnamefont {R.}~\bibnamefont {Reuter}}, \bibinfo
  {author} {\bibfnamefont {R.}~\bibnamefont {St{\"{o}}hr}}, \bibinfo {author}
  {\bibfnamefont {A.}~\bibnamefont {Zappe}}, \bibinfo {author} {\bibfnamefont
  {J.}~\bibnamefont {Meijer}}, \bibinfo {author} {\bibfnamefont
  {P.}~\bibnamefont {Hemmer}}, \ and\ \bibinfo {author} {\bibfnamefont
  {J.}~\bibnamefont {Wrachtrup}},\ }\href {\doibase 10.1038/ncomms2034}
  {\bibfield  {journal} {\bibinfo  {journal} {Nature Communications}\ }\textbf
  {\bibinfo {volume} {3}},\ \bibinfo {pages} {1029} (\bibinfo {year}
  {2012})}\BibitemShut {NoStop}%
\bibitem [{\citenamefont {Nakamura}\ \emph {et~al.}(2015)\citenamefont
  {Nakamura}, \citenamefont {Yoshihiro}, \citenamefont {Inagawa}, \citenamefont
  {Fujiyoshi},\ and\ \citenamefont {Matsushita}}]{Nakamura2014}%
  \BibitemOpen
  \bibfield  {author} {\bibinfo {author} {\bibfnamefont {I.}~\bibnamefont
  {Nakamura}}, \bibinfo {author} {\bibfnamefont {T.}~\bibnamefont {Yoshihiro}},
  \bibinfo {author} {\bibfnamefont {H.}~\bibnamefont {Inagawa}}, \bibinfo
  {author} {\bibfnamefont {S.}~\bibnamefont {Fujiyoshi}}, \ and\ \bibinfo
  {author} {\bibfnamefont {M.}~\bibnamefont {Matsushita}},\ }\href {\doibase
  10.1038/srep07364} {\bibfield  {journal} {\bibinfo  {journal} {Scientific
  Reports}\ }\textbf {\bibinfo {volume} {4}},\ \bibinfo {pages} {7364}
  (\bibinfo {year} {2015})}\BibitemShut {NoStop}%
\bibitem [{\citenamefont {Eichhammer}\ \emph {et~al.}(2015)\citenamefont
  {Eichhammer}, \citenamefont {Utikal}, \citenamefont {G{\"{o}}tzinger},\ and\
  \citenamefont {Sandoghdar}}]{Eichhammer2015}%
  \BibitemOpen
  \bibfield  {author} {\bibinfo {author} {\bibfnamefont {E.}~\bibnamefont
  {Eichhammer}}, \bibinfo {author} {\bibfnamefont {T.}~\bibnamefont {Utikal}},
  \bibinfo {author} {\bibfnamefont {S.}~\bibnamefont {G{\"{o}}tzinger}}, \ and\
  \bibinfo {author} {\bibfnamefont {V.}~\bibnamefont {Sandoghdar}},\ }\href
  {\doibase 10.1088/1367-2630/17/8/083018} {\bibfield  {journal} {\bibinfo
  {journal} {New Journal of Physics}\ }\textbf {\bibinfo {volume} {17}},\
  \bibinfo {pages} {083018} (\bibinfo {year} {2015})}\BibitemShut {NoStop}%
\bibitem [{\citenamefont {Zhong}\ \emph {et~al.}(2018)\citenamefont {Zhong},
  \citenamefont {Kindem}, \citenamefont {Bartholomew}, \citenamefont {Rochman},
  \citenamefont {Craiciu}, \citenamefont {Verma}, \citenamefont {Nam},
  \citenamefont {Marsili}, \citenamefont {Shaw}, \citenamefont {Beyer},\ and\
  \citenamefont {Faraon}}]{Zhong2018}%
  \BibitemOpen
  \bibfield  {author} {\bibinfo {author} {\bibfnamefont {T.}~\bibnamefont
  {Zhong}}, \bibinfo {author} {\bibfnamefont {J.~M.}\ \bibnamefont {Kindem}},
  \bibinfo {author} {\bibfnamefont {J.~G.}\ \bibnamefont {Bartholomew}},
  \bibinfo {author} {\bibfnamefont {J.}~\bibnamefont {Rochman}}, \bibinfo
  {author} {\bibfnamefont {I.}~\bibnamefont {Craiciu}}, \bibinfo {author}
  {\bibfnamefont {V.}~\bibnamefont {Verma}}, \bibinfo {author} {\bibfnamefont
  {S.~W.}\ \bibnamefont {Nam}}, \bibinfo {author} {\bibfnamefont
  {F.}~\bibnamefont {Marsili}}, \bibinfo {author} {\bibfnamefont {M.~D.}\
  \bibnamefont {Shaw}}, \bibinfo {author} {\bibfnamefont {A.~D.}\ \bibnamefont
  {Beyer}}, \ and\ \bibinfo {author} {\bibfnamefont {A.}~\bibnamefont
  {Faraon}},\ }\href {\doibase 10.1103/PhysRevLett.121.183603} {\bibfield
  {journal} {\bibinfo  {journal} {Physical Review Letters}\ }\textbf {\bibinfo
  {volume} {121}},\ \bibinfo {pages} {183603} (\bibinfo {year}
  {2018})}\BibitemShut {NoStop}%
\bibitem [{\citenamefont {Raha}\ \emph {et~al.}(2019)\citenamefont {Raha},
  \citenamefont {Chen}, \citenamefont {Phenicie}, \citenamefont {Ourari},
  \citenamefont {Dibos},\ and\ \citenamefont {Thompson}}]{Raha2019}%
  \BibitemOpen
  \bibfield  {author} {\bibinfo {author} {\bibfnamefont {M.}~\bibnamefont
  {Raha}}, \bibinfo {author} {\bibfnamefont {S.}~\bibnamefont {Chen}}, \bibinfo
  {author} {\bibfnamefont {C.~M.}\ \bibnamefont {Phenicie}}, \bibinfo {author}
  {\bibfnamefont {S.}~\bibnamefont {Ourari}}, \bibinfo {author} {\bibfnamefont
  {A.~M.}\ \bibnamefont {Dibos}}, \ and\ \bibinfo {author} {\bibfnamefont
  {J.~D.}\ \bibnamefont {Thompson}},\ }\href {http://arxiv.org/abs/1907.09992}
  {\bibfield  {journal} {\bibinfo  {journal} {eprint arXiv:1907.09992}\ }
  (\bibinfo {year} {2019})}\BibitemShut {NoStop}%
\bibitem [{\citenamefont {Kindem}\ \emph {et~al.}(2019)\citenamefont {Kindem},
  \citenamefont {Ruskuc}, \citenamefont {Bartholomew}, \citenamefont {Rochman},
  \citenamefont {Huan},\ and\ \citenamefont {Faraon}}]{Kindem2019}%
  \BibitemOpen
  \bibfield  {author} {\bibinfo {author} {\bibfnamefont {J.~M.}\ \bibnamefont
  {Kindem}}, \bibinfo {author} {\bibfnamefont {A.}~\bibnamefont {Ruskuc}},
  \bibinfo {author} {\bibfnamefont {J.~G.}\ \bibnamefont {Bartholomew}},
  \bibinfo {author} {\bibfnamefont {J.}~\bibnamefont {Rochman}}, \bibinfo
  {author} {\bibfnamefont {Y.~Q.}\ \bibnamefont {Huan}}, \ and\ \bibinfo
  {author} {\bibfnamefont {A.}~\bibnamefont {Faraon}},\ }\href
  {http://arxiv.org/abs/1907.12161} {\bibfield  {journal} {\bibinfo  {journal}
  {eprint arXiv:1907.12161}\ } (\bibinfo {year} {2019})}\BibitemShut {NoStop}%
\bibitem [{\citenamefont {Williamson}\ \emph {et~al.}(2014)\citenamefont
  {Williamson}, \citenamefont {Chen},\ and\ \citenamefont
  {Longdell}}]{Williamson2014}%
  \BibitemOpen
  \bibfield  {author} {\bibinfo {author} {\bibfnamefont {L.~A.}\ \bibnamefont
  {Williamson}}, \bibinfo {author} {\bibfnamefont {Y.~H.}\ \bibnamefont
  {Chen}}, \ and\ \bibinfo {author} {\bibfnamefont {J.~J.}\ \bibnamefont
  {Longdell}},\ }\href {\doibase 10.1103/PhysRevLett.113.203601} {\bibfield
  {journal} {\bibinfo  {journal} {Physical Review Letters}\ }\textbf {\bibinfo
  {volume} {113}},\ \bibinfo {pages} {203601} (\bibinfo {year}
  {2014})}\BibitemShut {NoStop}%
\bibitem [{\citenamefont {O'Brien}\ \emph {et~al.}(2014)\citenamefont
  {O'Brien}, \citenamefont {Lauk}, \citenamefont {Blum}, \citenamefont
  {Morigi},\ and\ \citenamefont {Fleischhauer}}]{OBrien2014}%
  \BibitemOpen
  \bibfield  {author} {\bibinfo {author} {\bibfnamefont {C.}~\bibnamefont
  {O'Brien}}, \bibinfo {author} {\bibfnamefont {N.}~\bibnamefont {Lauk}},
  \bibinfo {author} {\bibfnamefont {S.}~\bibnamefont {Blum}}, \bibinfo {author}
  {\bibfnamefont {G.}~\bibnamefont {Morigi}}, \ and\ \bibinfo {author}
  {\bibfnamefont {M.}~\bibnamefont {Fleischhauer}},\ }\href {\doibase
  10.1103/PhysRevLett.113.063603} {\bibfield  {journal} {\bibinfo  {journal}
  {Physical Review Letters}\ }\textbf {\bibinfo {volume} {113}},\ \bibinfo
  {pages} {063603} (\bibinfo {year} {2014})}\BibitemShut {NoStop}%
\bibitem [{\citenamefont {Fernandez-Gonzalvo}\ \emph
  {et~al.}(2015)\citenamefont {Fernandez-Gonzalvo}, \citenamefont {Chen},
  \citenamefont {Yin}, \citenamefont {Rogge},\ and\ \citenamefont
  {Longdell}}]{Fernandez-Gonzalvo2015}%
  \BibitemOpen
  \bibfield  {author} {\bibinfo {author} {\bibfnamefont {X.}~\bibnamefont
  {Fernandez-Gonzalvo}}, \bibinfo {author} {\bibfnamefont {Y.-H.}\ \bibnamefont
  {Chen}}, \bibinfo {author} {\bibfnamefont {C.}~\bibnamefont {Yin}}, \bibinfo
  {author} {\bibfnamefont {S.}~\bibnamefont {Rogge}}, \ and\ \bibinfo {author}
  {\bibfnamefont {J.~J.}\ \bibnamefont {Longdell}},\ }\href {\doibase
  10.1103/PhysRevA.92.062313} {\bibfield  {journal} {\bibinfo  {journal}
  {Physical Review A}\ }\textbf {\bibinfo {volume} {92}},\ \bibinfo {pages}
  {062313} (\bibinfo {year} {2015})}\BibitemShut {NoStop}%
\bibitem [{\citenamefont {Dibos}\ \emph {et~al.}(2018)\citenamefont {Dibos},
  \citenamefont {Raha}, \citenamefont {Phenicie},\ and\ \citenamefont
  {Thompson}}]{Dibos2018}%
  \BibitemOpen
  \bibfield  {author} {\bibinfo {author} {\bibfnamefont {A.~M.}\ \bibnamefont
  {Dibos}}, \bibinfo {author} {\bibfnamefont {M.}~\bibnamefont {Raha}},
  \bibinfo {author} {\bibfnamefont {C.~M.}\ \bibnamefont {Phenicie}}, \ and\
  \bibinfo {author} {\bibfnamefont {J.~D.}\ \bibnamefont {Thompson}},\ }\href
  {\doibase 10.1103/PhysRevLett.120.243601} {\bibfield  {journal} {\bibinfo
  {journal} {Physical Review Letters}\ }\textbf {\bibinfo {volume} {120}},\
  \bibinfo {pages} {243601} (\bibinfo {year} {2018})}\BibitemShut {NoStop}%
\bibitem [{\citenamefont {McAuslan}\ \emph {et~al.}(2012)\citenamefont
  {McAuslan}, \citenamefont {Bartholomew}, \citenamefont {Sellars},\ and\
  \citenamefont {Longdell}}]{McAuslan2012}%
  \BibitemOpen
  \bibfield  {author} {\bibinfo {author} {\bibfnamefont {D.~L.}\ \bibnamefont
  {McAuslan}}, \bibinfo {author} {\bibfnamefont {J.~G.}\ \bibnamefont
  {Bartholomew}}, \bibinfo {author} {\bibfnamefont {M.~J.}\ \bibnamefont
  {Sellars}}, \ and\ \bibinfo {author} {\bibfnamefont {J.~J.}\ \bibnamefont
  {Longdell}},\ }\href {\doibase 10.1103/PhysRevA.85.032339} {\bibfield
  {journal} {\bibinfo  {journal} {Physical Review A}\ }\textbf {\bibinfo
  {volume} {85}},\ \bibinfo {pages} {032339} (\bibinfo {year}
  {2012})}\BibitemShut {NoStop}%
\bibitem [{\citenamefont {Ortu}\ \emph {et~al.}(2018)\citenamefont {Ortu},
  \citenamefont {Tiranov}, \citenamefont {Welinski}, \citenamefont
  {Fr{\"{o}}wis}, \citenamefont {Gisin}, \citenamefont {Ferrier}, \citenamefont
  {Goldner},\ and\ \citenamefont {Afzelius}}]{Ortu2018}%
  \BibitemOpen
  \bibfield  {author} {\bibinfo {author} {\bibfnamefont {A.}~\bibnamefont
  {Ortu}}, \bibinfo {author} {\bibfnamefont {A.}~\bibnamefont {Tiranov}},
  \bibinfo {author} {\bibfnamefont {S.}~\bibnamefont {Welinski}}, \bibinfo
  {author} {\bibfnamefont {F.}~\bibnamefont {Fr{\"{o}}wis}}, \bibinfo {author}
  {\bibfnamefont {N.}~\bibnamefont {Gisin}}, \bibinfo {author} {\bibfnamefont
  {A.}~\bibnamefont {Ferrier}}, \bibinfo {author} {\bibfnamefont
  {P.}~\bibnamefont {Goldner}}, \ and\ \bibinfo {author} {\bibfnamefont
  {M.}~\bibnamefont {Afzelius}},\ }\href {\doibase 10.1038/s41563-018-0138-x}
  {\bibfield  {journal} {\bibinfo  {journal} {Nature Materials}\ }\textbf
  {\bibinfo {volume} {17}},\ \bibinfo {pages} {671} (\bibinfo {year}
  {2018})}\BibitemShut {NoStop}%
\bibitem [{\citenamefont {Lim}\ \emph {et~al.}(2018)\citenamefont {Lim},
  \citenamefont {Welinski}, \citenamefont {Ferrier}, \citenamefont {Goldner},\
  and\ \citenamefont {Morton}}]{Lim2018}%
  \BibitemOpen
  \bibfield  {author} {\bibinfo {author} {\bibfnamefont {H.~J.}\ \bibnamefont
  {Lim}}, \bibinfo {author} {\bibfnamefont {S.}~\bibnamefont {Welinski}},
  \bibinfo {author} {\bibfnamefont {A.}~\bibnamefont {Ferrier}}, \bibinfo
  {author} {\bibfnamefont {P.}~\bibnamefont {Goldner}}, \ and\ \bibinfo
  {author} {\bibfnamefont {J.~J.}\ \bibnamefont {Morton}},\ }\href {\doibase
  10.1103/PhysRevB.97.064409} {\bibfield  {journal} {\bibinfo  {journal}
  {Physical Review B}\ }\textbf {\bibinfo {volume} {97}},\ \bibinfo {pages}
  {064409} (\bibinfo {year} {2018})}\BibitemShut {NoStop}%
\bibitem [{\citenamefont {Rakhmatullin}\ \emph {et~al.}(2009)\citenamefont
  {Rakhmatullin}, \citenamefont {Kurkin}, \citenamefont {Mamin}, \citenamefont
  {Orlinskii}, \citenamefont {Gafurov}, \citenamefont {Baibekov}, \citenamefont
  {Malkin}, \citenamefont {Gambarelli}, \citenamefont {Bertaina},\ and\
  \citenamefont {Barbara}}]{Rakhmatullin2009}%
  \BibitemOpen
  \bibfield  {author} {\bibinfo {author} {\bibfnamefont {R.~M.}\ \bibnamefont
  {Rakhmatullin}}, \bibinfo {author} {\bibfnamefont {I.~N.}\ \bibnamefont
  {Kurkin}}, \bibinfo {author} {\bibfnamefont {G.~V.}\ \bibnamefont {Mamin}},
  \bibinfo {author} {\bibfnamefont {S.~B.}\ \bibnamefont {Orlinskii}}, \bibinfo
  {author} {\bibfnamefont {M.~R.}\ \bibnamefont {Gafurov}}, \bibinfo {author}
  {\bibfnamefont {E.~I.}\ \bibnamefont {Baibekov}}, \bibinfo {author}
  {\bibfnamefont {B.~Z.}\ \bibnamefont {Malkin}}, \bibinfo {author}
  {\bibfnamefont {S.}~\bibnamefont {Gambarelli}}, \bibinfo {author}
  {\bibfnamefont {S.}~\bibnamefont {Bertaina}}, \ and\ \bibinfo {author}
  {\bibfnamefont {B.}~\bibnamefont {Barbara}},\ }\href {\doibase
  10.1103/PhysRevB.79.172408} {\bibfield  {journal} {\bibinfo  {journal}
  {Physical Review B}\ }\textbf {\bibinfo {volume} {79}},\ \bibinfo {pages}
  {172408} (\bibinfo {year} {2009})}\BibitemShut {NoStop}%
\bibitem [{\citenamefont {Liu}\ and\ \citenamefont {Jacquier}(2005)}]{Liu2005}%
  \BibitemOpen
  \bibfield  {author} {\bibinfo {author} {\bibfnamefont {G.}~\bibnamefont
  {Liu}}\ and\ \bibinfo {author} {\bibfnamefont {B.}~\bibnamefont {Jacquier}},\
  }\href {\doibase 10.1007/3-540-28209-2} {\emph {\bibinfo {title}
  {{Spectroscopic Properties of Rare Earths in Optical Materials}}}}\ (\bibinfo
   {publisher} {Springer-Verlag Berlin Heidelberg},\ \bibinfo {year}
  {2005})\BibitemShut {NoStop}%
\bibitem [{\citenamefont {Toyli}\ \emph {et~al.}(2010)\citenamefont {Toyli},
  \citenamefont {Weis}, \citenamefont {Fuchs}, \citenamefont {Schenkel},\ and\
  \citenamefont {Awschalom}}]{Toyli2010}%
  \BibitemOpen
  \bibfield  {author} {\bibinfo {author} {\bibfnamefont {D.~M.}\ \bibnamefont
  {Toyli}}, \bibinfo {author} {\bibfnamefont {C.~D.}\ \bibnamefont {Weis}},
  \bibinfo {author} {\bibfnamefont {G.~D.}\ \bibnamefont {Fuchs}}, \bibinfo
  {author} {\bibfnamefont {T.}~\bibnamefont {Schenkel}}, \ and\ \bibinfo
  {author} {\bibfnamefont {D.~D.}\ \bibnamefont {Awschalom}},\ }\href {\doibase
  10.1021/nl102066q} {\bibfield  {journal} {\bibinfo  {journal} {Nano Letters}\
  }\textbf {\bibinfo {volume} {10}},\ \bibinfo {pages} {3168} (\bibinfo {year}
  {2010})}\BibitemShut {NoStop}%
\bibitem [{\citenamefont {Chu}\ \emph {et~al.}(2014)\citenamefont {Chu},
  \citenamefont {{De Leon}}, \citenamefont {Shields}, \citenamefont {Hausmann},
  \citenamefont {Evans}, \citenamefont {Togan}, \citenamefont {Burek},
  \citenamefont {Markham}, \citenamefont {Stacey}, \citenamefont {Zibrov},
  \citenamefont {Yacoby}, \citenamefont {Twitchen}, \citenamefont {Loncar},
  \citenamefont {Park}, \citenamefont {Maletinsky},\ and\ \citenamefont
  {Lukin}}]{Chu2014}%
  \BibitemOpen
  \bibfield  {author} {\bibinfo {author} {\bibfnamefont {Y.}~\bibnamefont
  {Chu}}, \bibinfo {author} {\bibfnamefont {N.~P.}\ \bibnamefont {{De Leon}}},
  \bibinfo {author} {\bibfnamefont {B.~J.}\ \bibnamefont {Shields}}, \bibinfo
  {author} {\bibfnamefont {B.}~\bibnamefont {Hausmann}}, \bibinfo {author}
  {\bibfnamefont {R.}~\bibnamefont {Evans}}, \bibinfo {author} {\bibfnamefont
  {E.}~\bibnamefont {Togan}}, \bibinfo {author} {\bibfnamefont {M.~J.}\
  \bibnamefont {Burek}}, \bibinfo {author} {\bibfnamefont {M.}~\bibnamefont
  {Markham}}, \bibinfo {author} {\bibfnamefont {A.}~\bibnamefont {Stacey}},
  \bibinfo {author} {\bibfnamefont {A.~S.}\ \bibnamefont {Zibrov}}, \bibinfo
  {author} {\bibfnamefont {A.}~\bibnamefont {Yacoby}}, \bibinfo {author}
  {\bibfnamefont {D.~J.}\ \bibnamefont {Twitchen}}, \bibinfo {author}
  {\bibfnamefont {M.}~\bibnamefont {Loncar}}, \bibinfo {author} {\bibfnamefont
  {H.}~\bibnamefont {Park}}, \bibinfo {author} {\bibfnamefont {P.}~\bibnamefont
  {Maletinsky}}, \ and\ \bibinfo {author} {\bibfnamefont {M.~D.}\ \bibnamefont
  {Lukin}},\ }\href {\doibase 10.1021/nl404836p} {\bibfield  {journal}
  {\bibinfo  {journal} {Nano Letters}\ }\textbf {\bibinfo {volume} {14}},\
  \bibinfo {pages} {1982} (\bibinfo {year} {2014})}\BibitemShut {NoStop}%
\bibitem [{\citenamefont {Rose}\ \emph {et~al.}(2018)\citenamefont {Rose},
  \citenamefont {Huang}, \citenamefont {Zhang}, \citenamefont {Stevenson},
  \citenamefont {Tyryshkin}, \citenamefont {Sangtawesin}, \citenamefont
  {Srinivasan}, \citenamefont {Loudin}, \citenamefont {Markham}, \citenamefont
  {Edmonds}, \citenamefont {Twitchen}, \citenamefont {Lyon},\ and\
  \citenamefont {de~Leon}}]{Rose2018}%
  \BibitemOpen
  \bibfield  {author} {\bibinfo {author} {\bibfnamefont {B.~C.}\ \bibnamefont
  {Rose}}, \bibinfo {author} {\bibfnamefont {D.}~\bibnamefont {Huang}},
  \bibinfo {author} {\bibfnamefont {Z.-H.}\ \bibnamefont {Zhang}}, \bibinfo
  {author} {\bibfnamefont {P.}~\bibnamefont {Stevenson}}, \bibinfo {author}
  {\bibfnamefont {A.~M.}\ \bibnamefont {Tyryshkin}}, \bibinfo {author}
  {\bibfnamefont {S.}~\bibnamefont {Sangtawesin}}, \bibinfo {author}
  {\bibfnamefont {S.}~\bibnamefont {Srinivasan}}, \bibinfo {author}
  {\bibfnamefont {L.}~\bibnamefont {Loudin}}, \bibinfo {author} {\bibfnamefont
  {M.~L.}\ \bibnamefont {Markham}}, \bibinfo {author} {\bibfnamefont {A.~M.}\
  \bibnamefont {Edmonds}}, \bibinfo {author} {\bibfnamefont {D.~J.}\
  \bibnamefont {Twitchen}}, \bibinfo {author} {\bibfnamefont {S.~A.}\
  \bibnamefont {Lyon}}, \ and\ \bibinfo {author} {\bibfnamefont {N.~P.}\
  \bibnamefont {de~Leon}},\ }\href {\doibase 10.1126/science.aao0290}
  {\bibfield  {journal} {\bibinfo  {journal} {Science}\ }\textbf {\bibinfo
  {volume} {361}},\ \bibinfo {pages} {60} (\bibinfo {year} {2018})}\BibitemShut
  {NoStop}%
\bibitem [{\citenamefont {Sipahigil}\ \emph {et~al.}(2016)\citenamefont
  {Sipahigil}, \citenamefont {Evans}, \citenamefont {Sukachev}, \citenamefont
  {Burek}, \citenamefont {Borregaard}, \citenamefont {Bhaskar}, \citenamefont
  {Nguyen}, \citenamefont {Pacheco}, \citenamefont {Atikian}, \citenamefont
  {Meuwly}, \citenamefont {Camacho}, \citenamefont {Jelezko}, \citenamefont
  {Bielejec}, \citenamefont {Park}, \citenamefont {Loncar},\ and\ \citenamefont
  {Lukin}}]{Sipahigil2016}%
  \BibitemOpen
  \bibfield  {author} {\bibinfo {author} {\bibfnamefont {A.}~\bibnamefont
  {Sipahigil}}, \bibinfo {author} {\bibfnamefont {R.~E.}\ \bibnamefont
  {Evans}}, \bibinfo {author} {\bibfnamefont {D.~D.}\ \bibnamefont {Sukachev}},
  \bibinfo {author} {\bibfnamefont {M.~J.}\ \bibnamefont {Burek}}, \bibinfo
  {author} {\bibfnamefont {J.}~\bibnamefont {Borregaard}}, \bibinfo {author}
  {\bibfnamefont {M.~K.}\ \bibnamefont {Bhaskar}}, \bibinfo {author}
  {\bibfnamefont {C.~T.}\ \bibnamefont {Nguyen}}, \bibinfo {author}
  {\bibfnamefont {J.~L.}\ \bibnamefont {Pacheco}}, \bibinfo {author}
  {\bibfnamefont {H.~A.}\ \bibnamefont {Atikian}}, \bibinfo {author}
  {\bibfnamefont {C.}~\bibnamefont {Meuwly}}, \bibinfo {author} {\bibfnamefont
  {R.~M.}\ \bibnamefont {Camacho}}, \bibinfo {author} {\bibfnamefont
  {F.}~\bibnamefont {Jelezko}}, \bibinfo {author} {\bibfnamefont
  {E.}~\bibnamefont {Bielejec}}, \bibinfo {author} {\bibfnamefont
  {H.}~\bibnamefont {Park}}, \bibinfo {author} {\bibfnamefont {M.}~\bibnamefont
  {Loncar}}, \ and\ \bibinfo {author} {\bibfnamefont {M.~D.}\ \bibnamefont
  {Lukin}},\ }\href {\doibase 10.1126/science.aah6875} {\bibfield  {journal}
  {\bibinfo  {journal} {Science}\ }\textbf {\bibinfo {volume} {354}},\ \bibinfo
  {pages} {847} (\bibinfo {year} {2016})}\BibitemShut {NoStop}%
\bibitem [{\citenamefont {Probst}\ \emph {et~al.}(2014)\citenamefont {Probst},
  \citenamefont {Kukharchyk}, \citenamefont {Rotzinger}, \citenamefont
  {Tkal{\v{c}}ec}, \citenamefont {W{\"{u}}nsch}, \citenamefont {Wieck},
  \citenamefont {Siegel}, \citenamefont {Ustinov},\ and\ \citenamefont
  {Bushev}}]{Probst2014}%
  \BibitemOpen
  \bibfield  {author} {\bibinfo {author} {\bibfnamefont {S.}~\bibnamefont
  {Probst}}, \bibinfo {author} {\bibfnamefont {N.}~\bibnamefont {Kukharchyk}},
  \bibinfo {author} {\bibfnamefont {H.}~\bibnamefont {Rotzinger}}, \bibinfo
  {author} {\bibfnamefont {A.}~\bibnamefont {Tkal{\v{c}}ec}}, \bibinfo {author}
  {\bibfnamefont {S.}~\bibnamefont {W{\"{u}}nsch}}, \bibinfo {author}
  {\bibfnamefont {A.~D.}\ \bibnamefont {Wieck}}, \bibinfo {author}
  {\bibfnamefont {M.}~\bibnamefont {Siegel}}, \bibinfo {author} {\bibfnamefont
  {A.~V.}\ \bibnamefont {Ustinov}}, \ and\ \bibinfo {author} {\bibfnamefont
  {P.~A.}\ \bibnamefont {Bushev}},\ }\href {\doibase 10.1063/1.4898696}
  {\bibfield  {journal} {\bibinfo  {journal} {Applied Physics Letters}\
  }\textbf {\bibinfo {volume} {105}},\ \bibinfo {pages} {162404} (\bibinfo
  {year} {2014})}\BibitemShut {NoStop}%
\bibitem [{\citenamefont {Wisby}\ \emph {et~al.}(2016)\citenamefont {Wisby},
  \citenamefont {{De Graaf}}, \citenamefont {Gwilliam}, \citenamefont
  {Adamyan}, \citenamefont {Kubatkin}, \citenamefont {Meeson}, \citenamefont
  {Tzalenchuk},\ and\ \citenamefont {Lindstr{\"{o}}m}}]{Wisby2016}%
  \BibitemOpen
  \bibfield  {author} {\bibinfo {author} {\bibfnamefont {I.~S.}\ \bibnamefont
  {Wisby}}, \bibinfo {author} {\bibfnamefont {S.~E.}\ \bibnamefont {{De
  Graaf}}}, \bibinfo {author} {\bibfnamefont {R.}~\bibnamefont {Gwilliam}},
  \bibinfo {author} {\bibfnamefont {A.}~\bibnamefont {Adamyan}}, \bibinfo
  {author} {\bibfnamefont {S.~E.}\ \bibnamefont {Kubatkin}}, \bibinfo {author}
  {\bibfnamefont {P.~J.}\ \bibnamefont {Meeson}}, \bibinfo {author}
  {\bibfnamefont {A.~Y.}\ \bibnamefont {Tzalenchuk}}, \ and\ \bibinfo {author}
  {\bibfnamefont {T.}~\bibnamefont {Lindstr{\"{o}}m}},\ }\href {\doibase
  10.1103/PhysRevApplied.6.024021} {\bibfield  {journal} {\bibinfo  {journal}
  {Physical Review Applied}\ }\textbf {\bibinfo {volume} {6}},\ \bibinfo
  {pages} {024021} (\bibinfo {year} {2016})}\BibitemShut {NoStop}%
\bibitem [{\citenamefont {Xia}\ \emph {et~al.}(2015)\citenamefont {Xia},
  \citenamefont {Kolesov}, \citenamefont {Wang}, \citenamefont {Siyushev},
  \citenamefont {Reuter}, \citenamefont {Kornher}, \citenamefont {Kukharchyk},
  \citenamefont {Wieck}, \citenamefont {Villa}, \citenamefont {Yang},\ and\
  \citenamefont {Wrachtrup}}]{Xia2015}%
  \BibitemOpen
  \bibfield  {author} {\bibinfo {author} {\bibfnamefont {K.}~\bibnamefont
  {Xia}}, \bibinfo {author} {\bibfnamefont {R.}~\bibnamefont {Kolesov}},
  \bibinfo {author} {\bibfnamefont {Y.}~\bibnamefont {Wang}}, \bibinfo {author}
  {\bibfnamefont {P.}~\bibnamefont {Siyushev}}, \bibinfo {author}
  {\bibfnamefont {R.}~\bibnamefont {Reuter}}, \bibinfo {author} {\bibfnamefont
  {T.}~\bibnamefont {Kornher}}, \bibinfo {author} {\bibfnamefont
  {N.}~\bibnamefont {Kukharchyk}}, \bibinfo {author} {\bibfnamefont {A.~D.}\
  \bibnamefont {Wieck}}, \bibinfo {author} {\bibfnamefont {B.}~\bibnamefont
  {Villa}}, \bibinfo {author} {\bibfnamefont {S.}~\bibnamefont {Yang}}, \ and\
  \bibinfo {author} {\bibfnamefont {J.}~\bibnamefont {Wrachtrup}},\ }\href
  {\doibase 10.1103/PhysRevLett.115.093602} {\bibfield  {journal} {\bibinfo
  {journal} {Physical Review Letters}\ }\textbf {\bibinfo {volume} {115}},\
  \bibinfo {pages} {093602} (\bibinfo {year} {2015})}\BibitemShut {NoStop}%
\bibitem [{\citenamefont {Kornher}\ \emph {et~al.}(2016)\citenamefont
  {Kornher}, \citenamefont {Xia}, \citenamefont {Kolesov}, \citenamefont
  {Kukharchyk}, \citenamefont {Reuter}, \citenamefont {Siyushev}, \citenamefont
  {St{\"{o}}hr}, \citenamefont {Schreck}, \citenamefont {Becker}, \citenamefont
  {Villa}, \citenamefont {Wieck},\ and\ \citenamefont
  {Wrachtrup}}]{Kornher2016}%
  \BibitemOpen
  \bibfield  {author} {\bibinfo {author} {\bibfnamefont {T.}~\bibnamefont
  {Kornher}}, \bibinfo {author} {\bibfnamefont {K.}~\bibnamefont {Xia}},
  \bibinfo {author} {\bibfnamefont {R.}~\bibnamefont {Kolesov}}, \bibinfo
  {author} {\bibfnamefont {N.}~\bibnamefont {Kukharchyk}}, \bibinfo {author}
  {\bibfnamefont {R.}~\bibnamefont {Reuter}}, \bibinfo {author} {\bibfnamefont
  {P.}~\bibnamefont {Siyushev}}, \bibinfo {author} {\bibfnamefont
  {R.}~\bibnamefont {St{\"{o}}hr}}, \bibinfo {author} {\bibfnamefont
  {M.}~\bibnamefont {Schreck}}, \bibinfo {author} {\bibfnamefont {H.~W.}\
  \bibnamefont {Becker}}, \bibinfo {author} {\bibfnamefont {B.}~\bibnamefont
  {Villa}}, \bibinfo {author} {\bibfnamefont {A.~D.}\ \bibnamefont {Wieck}}, \
  and\ \bibinfo {author} {\bibfnamefont {J.}~\bibnamefont {Wrachtrup}},\ }\href
  {\doibase 10.1063/1.4941403} {\bibfield  {journal} {\bibinfo  {journal}
  {Applied Physics Letters}\ }\textbf {\bibinfo {volume} {108}},\ \bibinfo
  {pages} {053108} (\bibinfo {year} {2016})}\BibitemShut {NoStop}%
\bibitem [{\citenamefont {Groot-Berning}\ \emph {et~al.}(2019)\citenamefont
  {Groot-Berning}, \citenamefont {Kornher}, \citenamefont {Jacob},
  \citenamefont {Stopp}, \citenamefont {Dawkins}, \citenamefont {Kolesov},
  \citenamefont {Wrachtrup}, \citenamefont {Singer},\ and\ \citenamefont
  {Schmidt-Kaler}}]{Groot-Berning2019}%
  \BibitemOpen
  \bibfield  {author} {\bibinfo {author} {\bibfnamefont {K.}~\bibnamefont
  {Groot-Berning}}, \bibinfo {author} {\bibfnamefont {T.}~\bibnamefont
  {Kornher}}, \bibinfo {author} {\bibfnamefont {G.}~\bibnamefont {Jacob}},
  \bibinfo {author} {\bibfnamefont {F.}~\bibnamefont {Stopp}}, \bibinfo
  {author} {\bibfnamefont {S.~T.}\ \bibnamefont {Dawkins}}, \bibinfo {author}
  {\bibfnamefont {R.}~\bibnamefont {Kolesov}}, \bibinfo {author} {\bibfnamefont
  {J.}~\bibnamefont {Wrachtrup}}, \bibinfo {author} {\bibfnamefont
  {K.}~\bibnamefont {Singer}}, \ and\ \bibinfo {author} {\bibfnamefont
  {F.}~\bibnamefont {Schmidt-Kaler}},\ }\href {\doibase
  10.1103/PhysRevLett.123.106802} {\bibfield  {journal} {\bibinfo  {journal}
  {Physical Review Letters}\ }\textbf {\bibinfo {volume} {123}},\ \bibinfo
  {pages} {106802} (\bibinfo {year} {2019})}\BibitemShut {NoStop}%
\bibitem [{\citenamefont {Thiel}\ \emph {et~al.}(2012)\citenamefont {Thiel},
  \citenamefont {Sun}, \citenamefont {Macfarlane}, \citenamefont
  {B{\"{o}}ttger},\ and\ \citenamefont {Cone}}]{Thiel2012}%
  \BibitemOpen
  \bibfield  {author} {\bibinfo {author} {\bibfnamefont {C.~W.}\ \bibnamefont
  {Thiel}}, \bibinfo {author} {\bibfnamefont {Y.}~\bibnamefont {Sun}}, \bibinfo
  {author} {\bibfnamefont {R.~M.}\ \bibnamefont {Macfarlane}}, \bibinfo
  {author} {\bibfnamefont {T.}~\bibnamefont {B{\"{o}}ttger}}, \ and\ \bibinfo
  {author} {\bibfnamefont {R.~L.}\ \bibnamefont {Cone}},\ }\href {\doibase
  10.1088/0953-4075/45/12/124013} {\bibfield  {journal} {\bibinfo  {journal}
  {Journal of Physics B: Atomic, Molecular and Optical Physics}\ }\textbf
  {\bibinfo {volume} {45}},\ \bibinfo {pages} {124013} (\bibinfo {year}
  {2012})}\BibitemShut {NoStop}%
\bibitem [{Ger()}]{Gerritsen1962}%
  \BibitemOpen
  \href@noop {} {}\bibinfo {note} {Gerritsen, H.J. Paramagnetic Resonance of
  Transition Metal Ions in Rutile (TiO$_2$). In \emph{Paramagnetic Resonance,
  Proceedings of the First International Conference Held in Jerusalem},
  Jerusalem, Israel, 1962; Low, W., Ed.; Academic Press, (1963) pp.
  3-12}\BibitemShut {NoStop}%
\bibitem [{Sab()}]{Sabisky1966}%
  \BibitemOpen
  \href@noop {} {}\bibinfo {note} {Sabisky, E. S.; Gerritsen, H. J. Doubly
  doped titanium dioxide maser element. U.S. Patent 3,275,558, September 27,
  1966}\BibitemShut {NoStop}%
\bibitem [{\citenamefont {Thiel}\ \emph {et~al.}(2011)\citenamefont {Thiel},
  \citenamefont {B{\"{o}}ttger},\ and\ \citenamefont {Cone}}]{Thiel2011}%
  \BibitemOpen
  \bibfield  {author} {\bibinfo {author} {\bibfnamefont {C.}~\bibnamefont
  {Thiel}}, \bibinfo {author} {\bibfnamefont {T.}~\bibnamefont
  {B{\"{o}}ttger}}, \ and\ \bibinfo {author} {\bibfnamefont {R.}~\bibnamefont
  {Cone}},\ }\href {\doibase 10.1016/j.jlumin.2010.12.015} {\bibfield
  {journal} {\bibinfo  {journal} {Journal of Luminescence}\ }\textbf {\bibinfo
  {volume} {131}},\ \bibinfo {pages} {353} (\bibinfo {year}
  {2011})}\BibitemShut {NoStop}%
\bibitem [{sup()}]{suppinfo}%
  \BibitemOpen
  \href@noop {} {}\bibinfo {note} {See Supporting Information}\BibitemShut
  {NoStop}%
\bibitem [{\citenamefont {B{\"{o}}ttger}\ \emph {et~al.}(2006)\citenamefont
  {B{\"{o}}ttger}, \citenamefont {Sun}, \citenamefont {Thiel},\ and\
  \citenamefont {Cone}}]{Bottger2006}%
  \BibitemOpen
  \bibfield  {author} {\bibinfo {author} {\bibfnamefont {T.}~\bibnamefont
  {B{\"{o}}ttger}}, \bibinfo {author} {\bibfnamefont {Y.}~\bibnamefont {Sun}},
  \bibinfo {author} {\bibfnamefont {C.~W.}\ \bibnamefont {Thiel}}, \ and\
  \bibinfo {author} {\bibfnamefont {R.~L.}\ \bibnamefont {Cone}},\ }\href
  {\doibase 10.1103/PhysRevB.74.075107} {\bibfield  {journal} {\bibinfo
  {journal} {Physical Review B}\ }\textbf {\bibinfo {volume} {74}},\ \bibinfo
  {pages} {075107} (\bibinfo {year} {2006})}\BibitemShut {NoStop}%
\bibitem [{\citenamefont {Carter}\ and\ \citenamefont
  {Okaya}(1960)}]{Carter1960}%
  \BibitemOpen
  \bibfield  {author} {\bibinfo {author} {\bibfnamefont {D.~L.}\ \bibnamefont
  {Carter}}\ and\ \bibinfo {author} {\bibfnamefont {A.}~\bibnamefont {Okaya}},\
  }\href {\doibase 10.1103/PhysRev.118.1485} {\bibfield  {journal} {\bibinfo
  {journal} {Physical Review}\ }\textbf {\bibinfo {volume} {118}},\ \bibinfo
  {pages} {1485} (\bibinfo {year} {1960})}\BibitemShut {NoStop}%
\bibitem [{\citenamefont {Gerritsen}\ and\ \citenamefont
  {Sabisky}(1963)}]{Gerritsen1963}%
  \BibitemOpen
  \bibfield  {author} {\bibinfo {author} {\bibfnamefont {H.~J.}\ \bibnamefont
  {Gerritsen}}\ and\ \bibinfo {author} {\bibfnamefont {E.~S.}\ \bibnamefont
  {Sabisky}},\ }\href {\doibase 10.1103/PhysRev.132.1507} {\bibfield  {journal}
  {\bibinfo  {journal} {Physical Review}\ }\textbf {\bibinfo {volume} {132}},\
  \bibinfo {pages} {1507} (\bibinfo {year} {1963})}\BibitemShut {NoStop}%
\bibitem [{\citenamefont {Gerritsen}\ and\ \citenamefont
  {Sabisky}(1962)}]{Gerritsen1962Ni}%
  \BibitemOpen
  \bibfield  {author} {\bibinfo {author} {\bibfnamefont {H.~J.}\ \bibnamefont
  {Gerritsen}}\ and\ \bibinfo {author} {\bibfnamefont {E.~S.}\ \bibnamefont
  {Sabisky}},\ }\href {\doibase 10.1103/PhysRev.125.1853} {\bibfield  {journal}
  {\bibinfo  {journal} {Physical Review}\ }\textbf {\bibinfo {volume} {125}},\
  \bibinfo {pages} {1853} (\bibinfo {year} {1962})}\BibitemShut {NoStop}%
\bibitem [{\citenamefont {Dodson}\ and\ \citenamefont
  {Zia}(2012)}]{Dodson2012}%
  \BibitemOpen
  \bibfield  {author} {\bibinfo {author} {\bibfnamefont {C.~M.}\ \bibnamefont
  {Dodson}}\ and\ \bibinfo {author} {\bibfnamefont {R.}~\bibnamefont {Zia}},\
  }\href {\doibase 10.1103/PhysRevB.86.125102} {\bibfield  {journal} {\bibinfo
  {journal} {Physical Review B}\ }\textbf {\bibinfo {volume} {86}},\ \bibinfo
  {pages} {125102} (\bibinfo {year} {2012})}\BibitemShut {NoStop}%
\bibitem [{\citenamefont {Rams}\ \emph {et~al.}(1997)\citenamefont {Rams},
  \citenamefont {Tejeda},\ and\ \citenamefont {Cabrera}}]{Rams1997}%
  \BibitemOpen
  \bibfield  {author} {\bibinfo {author} {\bibfnamefont {J.}~\bibnamefont
  {Rams}}, \bibinfo {author} {\bibfnamefont {A.}~\bibnamefont {Tejeda}}, \ and\
  \bibinfo {author} {\bibfnamefont {J.~M.}\ \bibnamefont {Cabrera}},\ }\href
  {\doibase 10.1063/1.365938} {\bibfield  {journal} {\bibinfo  {journal}
  {Journal of Applied Physics}\ }\textbf {\bibinfo {volume} {82}},\ \bibinfo
  {pages} {994} (\bibinfo {year} {1997})}\BibitemShut {NoStop}%
\bibitem [{\citenamefont {Fromknecht}\ \emph {et~al.}(1996)\citenamefont
  {Fromknecht}, \citenamefont {Khubeis},\ and\ \citenamefont
  {Meyer}}]{Fromknecht1996}%
  \BibitemOpen
  \bibfield  {author} {\bibinfo {author} {\bibfnamefont {R.}~\bibnamefont
  {Fromknecht}}, \bibinfo {author} {\bibfnamefont {I.}~\bibnamefont {Khubeis}},
  \ and\ \bibinfo {author} {\bibfnamefont {O.}~\bibnamefont {Meyer}},\ }\href
  {\doibase 10.1016/0168-583X(96)00018-3} {\bibfield  {journal} {\bibinfo
  {journal} {Nuclear Instruments and Methods in Physics Research Section B:
  Beam Interactions with Materials and Atoms}\ }\textbf {\bibinfo {volume}
  {116}},\ \bibinfo {pages} {109} (\bibinfo {year} {1996})}\BibitemShut
  {NoStop}%
\bibitem [{\citenamefont {Watkins}(1959)}]{Watkins1959}%
  \BibitemOpen
  \bibfield  {author} {\bibinfo {author} {\bibfnamefont {G.~D.}\ \bibnamefont
  {Watkins}},\ }\href {\doibase 10.1103/PhysRev.113.79} {\bibfield  {journal}
  {\bibinfo  {journal} {Physical Review}\ }\textbf {\bibinfo {volume} {113}},\
  \bibinfo {pages} {79} (\bibinfo {year} {1959})}\BibitemShut {NoStop}%
\bibitem [{\citenamefont {Mims}(1968)}]{Mims1968}%
  \BibitemOpen
  \bibfield  {author} {\bibinfo {author} {\bibfnamefont {W.~B.}\ \bibnamefont
  {Mims}},\ }\href {\doibase 10.1103/PhysRev.168.370} {\bibfield  {journal}
  {\bibinfo  {journal} {Physical Review}\ }\textbf {\bibinfo {volume} {168}},\
  \bibinfo {pages} {370} (\bibinfo {year} {1968})}\BibitemShut {NoStop}%
\bibitem [{\citenamefont {Cockroft}\ and\ \citenamefont
  {Wright}(1992)}]{Cockroft1992}%
  \BibitemOpen
  \bibfield  {author} {\bibinfo {author} {\bibfnamefont {N.~J.}\ \bibnamefont
  {Cockroft}}\ and\ \bibinfo {author} {\bibfnamefont {J.~C.}\ \bibnamefont
  {Wright}},\ }\href {\doibase 10.1103/PhysRevB.45.9642} {\bibfield  {journal}
  {\bibinfo  {journal} {Physical Review B}\ }\textbf {\bibinfo {volume} {45}},\
  \bibinfo {pages} {9642} (\bibinfo {year} {1992})}\BibitemShut {NoStop}%
\bibitem [{\citenamefont {Jiang}\ \emph {et~al.}(2009)\citenamefont {Jiang},
  \citenamefont {Hodges}, \citenamefont {Maze}, \citenamefont {Maurer},
  \citenamefont {Taylor}, \citenamefont {Cory}, \citenamefont {Hemmer},
  \citenamefont {Walsworth}, \citenamefont {Yacoby}, \citenamefont {Zibrov},\
  and\ \citenamefont {Lukin}}]{Jiang2009}%
  \BibitemOpen
  \bibfield  {author} {\bibinfo {author} {\bibfnamefont {L.}~\bibnamefont
  {Jiang}}, \bibinfo {author} {\bibfnamefont {J.~S.}\ \bibnamefont {Hodges}},
  \bibinfo {author} {\bibfnamefont {J.~R.}\ \bibnamefont {Maze}}, \bibinfo
  {author} {\bibfnamefont {P.}~\bibnamefont {Maurer}}, \bibinfo {author}
  {\bibfnamefont {J.~M.}\ \bibnamefont {Taylor}}, \bibinfo {author}
  {\bibfnamefont {D.~G.}\ \bibnamefont {Cory}}, \bibinfo {author}
  {\bibfnamefont {P.~R.}\ \bibnamefont {Hemmer}}, \bibinfo {author}
  {\bibfnamefont {R.~L.}\ \bibnamefont {Walsworth}}, \bibinfo {author}
  {\bibfnamefont {A.}~\bibnamefont {Yacoby}}, \bibinfo {author} {\bibfnamefont
  {A.~S.}\ \bibnamefont {Zibrov}}, \ and\ \bibinfo {author} {\bibfnamefont
  {M.~D.}\ \bibnamefont {Lukin}},\ }\href {\doibase 10.1126/science.1176496}
  {\bibfield  {journal} {\bibinfo  {journal} {Science}\ }\textbf {\bibinfo
  {volume} {326}},\ \bibinfo {pages} {267} (\bibinfo {year}
  {2009})}\BibitemShut {NoStop}%
\bibitem [{\citenamefont {Ghoshtagore}\ and\ \citenamefont
  {Noreika}(1970)}]{Ghoshtagore1970}%
  \BibitemOpen
  \bibfield  {author} {\bibinfo {author} {\bibfnamefont {R.~N.}\ \bibnamefont
  {Ghoshtagore}}\ and\ \bibinfo {author} {\bibfnamefont {A.~J.}\ \bibnamefont
  {Noreika}},\ }\href {\doibase 10.1149/1.2407298} {\bibfield  {journal}
  {\bibinfo  {journal} {Journal of the Electrochemical Society}\ }\textbf
  {\bibinfo {volume} {117}},\ \bibinfo {pages} {1310} (\bibinfo {year}
  {1970})}\BibitemShut {NoStop}%
\end{thebibliography}%


%merlin.mbs apsrev4-1.bst 2010-07-25 4.21a (PWD, AO, DPC) hacked
%Control: key (0)
%Control: author (72) initials jnrlst
%Control: editor formatted (1) identically to author
%Control: production of article title (-1) disabled
%Control: page (0) single
%Control: year (1) truncated
%Control: production of eprint (0) enabled
\begin{thebibliography}{6}%
\makeatletter
\providecommand \@ifxundefined [1]{%
 \@ifx{#1\undefined}
}%
\providecommand \@ifnum [1]{%
 \ifnum #1\expandafter \@firstoftwo
 \else \expandafter \@secondoftwo
 \fi
}%
\providecommand \@ifx [1]{%
 \ifx #1\expandafter \@firstoftwo
 \else \expandafter \@secondoftwo
 \fi
}%
\providecommand \natexlab [1]{#1}%
\providecommand \enquote  [1]{``#1''}%
\providecommand \bibnamefont  [1]{#1}%
\providecommand \bibfnamefont [1]{#1}%
\providecommand \citenamefont [1]{#1}%
\providecommand \href@noop [0]{\@secondoftwo}%
\providecommand \href [0]{\begingroup \@sanitize@url \@href}%
\providecommand \@href[1]{\@@startlink{#1}\@@href}%
\providecommand \@@href[1]{\endgroup#1\@@endlink}%
\providecommand \@sanitize@url [0]{\catcode `\\12\catcode `\$12\catcode
  `\&12\catcode `\#12\catcode `\^12\catcode `\_12\catcode `\%12\relax}%
\providecommand \@@startlink[1]{}%
\providecommand \@@endlink[0]{}%
\providecommand \url  [0]{\begingroup\@sanitize@url \@url }%
\providecommand \@url [1]{\endgroup\@href {#1}{\urlprefix }}%
\providecommand \urlprefix  [0]{URL }%
\providecommand \Eprint [0]{\href }%
\providecommand \doibase [0]{http://dx.doi.org/}%
\providecommand \selectlanguage [0]{\@gobble}%
\providecommand \bibinfo  [0]{\@secondoftwo}%
\providecommand \bibfield  [0]{\@secondoftwo}%
\providecommand \translation [1]{[#1]}%
\providecommand \BibitemOpen [0]{}%
\providecommand \bibitemStop [0]{}%
\providecommand \bibitemNoStop [0]{.\EOS\space}%
\providecommand \EOS [0]{\spacefactor3000\relax}%
\providecommand \BibitemShut  [1]{\csname bibitem#1\endcsname}%
\let\auto@bib@innerbib\@empty
%</preamble>
\bibitem [{\citenamefont {Ziegler}\ \emph {et~al.}(2010)\citenamefont
  {Ziegler}, \citenamefont {Ziegler},\ and\ \citenamefont
  {Biersack}}]{Ziegler2010}%
  \BibitemOpen
  \bibfield  {author} {\bibinfo {author} {\bibfnamefont {J.~F.}\ \bibnamefont
  {Ziegler}}, \bibinfo {author} {\bibfnamefont {M.}~\bibnamefont {Ziegler}}, \
  and\ \bibinfo {author} {\bibfnamefont {J.}~\bibnamefont {Biersack}},\ }\href
  {\doibase 10.1016/j.nimb.2010.02.091} {\bibfield  {journal} {\bibinfo
  {journal} {Nuclear Instruments and Methods in Physics Research Section B:
  Beam Interactions with Materials and Atoms}\ }\textbf {\bibinfo {volume}
  {268}},\ \bibinfo {pages} {1818} (\bibinfo {year} {2010})}\BibitemShut
  {NoStop}%
\bibitem [{\citenamefont {Khomenko}\ \emph {et~al.}(1998)\citenamefont
  {Khomenko}, \citenamefont {Langer}, \citenamefont {Rager},\ and\
  \citenamefont {Fett}}]{Khomenko1998}%
  \BibitemOpen
  \bibfield  {author} {\bibinfo {author} {\bibfnamefont {V.~M.}\ \bibnamefont
  {Khomenko}}, \bibinfo {author} {\bibfnamefont {K.}~\bibnamefont {Langer}},
  \bibinfo {author} {\bibfnamefont {H.}~\bibnamefont {Rager}}, \ and\ \bibinfo
  {author} {\bibfnamefont {A.}~\bibnamefont {Fett}},\ }\href {\doibase
  10.1007/s002690050124} {\bibfield  {journal} {\bibinfo  {journal} {Physics
  and Chemistry of Minerals}\ }\textbf {\bibinfo {volume} {25}},\ \bibinfo
  {pages} {338} (\bibinfo {year} {1998})}\BibitemShut {NoStop}%
\bibitem [{\citenamefont {Barr}\ \emph {et~al.}(2008)\citenamefont {Barr},
  \citenamefont {Eaton},\ and\ \citenamefont {Eaton}}]{Barr2008}%
  \BibitemOpen
  \bibfield  {author} {\bibinfo {author} {\bibfnamefont {D.}~\bibnamefont
  {Barr}}, \bibinfo {author} {\bibfnamefont {S.~S.}\ \bibnamefont {Eaton}}, \
  and\ \bibinfo {author} {\bibfnamefont {G.~R.}\ \bibnamefont {Eaton}},\ }in\
  \href
  {https://www.google.com/url?sa=t{\&}rct=j{\&}q={\&}esrc=s{\&}source=web{\&}cd=1{\&}ved=2ahUKEwj{\_}juWCx5XkAhVKqlkKHbYhAAkQFjAAegQIBBAC{\&}url=https{\%}3A{\%}2F{\%}2Fportfolio.du.edu{\%}2FdownloadItem{\%}2F220703{\&}usg=AOvVaw3{\_}x7GNTUafkTVEpFDKUhxH}
  {\emph {\bibinfo {booktitle} {31st Annual International EPR Symposium,
  Breckenridge, Colorado}}}\ (\bibinfo {year} {2008})\ pp.\ \bibinfo {pages}
  {1--125}\BibitemShut {NoStop}%
\bibitem [{\citenamefont {Sun}\ \emph {et~al.}(2008)\citenamefont {Sun},
  \citenamefont {B{\"{o}}ttger}, \citenamefont {Thiel},\ and\ \citenamefont
  {Cone}}]{Sun2008}%
  \BibitemOpen
  \bibfield  {author} {\bibinfo {author} {\bibfnamefont {Y.}~\bibnamefont
  {Sun}}, \bibinfo {author} {\bibfnamefont {T.}~\bibnamefont {B{\"{o}}ttger}},
  \bibinfo {author} {\bibfnamefont {C.~W.}\ \bibnamefont {Thiel}}, \ and\
  \bibinfo {author} {\bibfnamefont {R.~L.}\ \bibnamefont {Cone}},\ }\href
  {\doibase 10.1103/PhysRevB.77.085124} {\bibfield  {journal} {\bibinfo
  {journal} {Physical Review B}\ }\textbf {\bibinfo {volume} {77}},\ \bibinfo
  {pages} {085124} (\bibinfo {year} {2008})}\BibitemShut {NoStop}%
\bibitem [{\citenamefont {B{\"{o}}ttger}\ \emph {et~al.}(2006)\citenamefont
  {B{\"{o}}ttger}, \citenamefont {Sun}, \citenamefont {Thiel},\ and\
  \citenamefont {Cone}}]{Bottger2006}%
  \BibitemOpen
  \bibfield  {author} {\bibinfo {author} {\bibfnamefont {T.}~\bibnamefont
  {B{\"{o}}ttger}}, \bibinfo {author} {\bibfnamefont {Y.}~\bibnamefont {Sun}},
  \bibinfo {author} {\bibfnamefont {C.~W.}\ \bibnamefont {Thiel}}, \ and\
  \bibinfo {author} {\bibfnamefont {R.~L.}\ \bibnamefont {Cone}},\ }\href
  {\doibase 10.1103/PhysRevB.74.075107} {\bibfield  {journal} {\bibinfo
  {journal} {Physical Review B}\ }\textbf {\bibinfo {volume} {74}},\ \bibinfo
  {pages} {075107} (\bibinfo {year} {2006})}\BibitemShut {NoStop}%
\bibitem [{\citenamefont {Abragam}\ and\ \citenamefont
  {Bleaney}(1970)}]{Abragam1970}%
  \BibitemOpen
  \bibfield  {author} {\bibinfo {author} {\bibfnamefont {A.}~\bibnamefont
  {Abragam}}\ and\ \bibinfo {author} {\bibfnamefont {B.}~\bibnamefont
  {Bleaney}},\ }\href@noop {} {\emph {\bibinfo {title} {Electron Paramagnetic
  Resonance of Transition Ions}}}\ (\bibinfo  {publisher} {Oxford University
  Press},\ \bibinfo {year} {1970})\BibitemShut {NoStop}%
\end{thebibliography}%

\end{document}

\title{Supporting Information for ``Narrow optical linewidths in erbium implanted in \tio{}"}

\author[1,3]{Christopher M.~Phenicie}
\author[1,3]{Paul Stevenson}
\author[1,3]{Sacha Welinski}
\author[1]{Brendon C.~Rose}
\author[1]{Abraham T.~Asfaw}
\author[2]{Robert J.~Cava}
\author[1]{Stephen A.~Lyon}
\author[1]{Nathalie P.~de Leon}
\author[1]{Jeff D. Thompson\thanks{jdthompson@princeton.edu}}

\affil[1]{Department of Electrical Engineering, Princeton University, Princeton, NJ 08544, USA}
\affil[2]{Department of Chemistry, Princeton University, Princeton, NJ 08544, USA}
\affil[3]{Contributed equally to this work}

\date{\today}
\maketitle

%%%%%%%%%%%%%%%%%%%%%%%%%%%%%%%%%
%%%%%%%%%%%%% SI %%%%%%%%%%%%%%%%

\renewcommand{\thetable}{S\arabic{table}} 
\setcounter{table}{0}
\renewcommand{\thefigure}{S\arabic{figure}} 
\setcounter{figure}{0}

\section{Experimental Details}

Rutile \tio{} samples were obtained from MTI (TOb101005S1) and implanted with doubly-ionized erbium (Innovion). The implantation was performed in multiple steps (Table \ref{tab:implantation}) to give a uniform density of implanted ions over a 100 nm depth, as calculated by Stopping-Range of Ions in Matter (SRIM) simulations~\cite{Ziegler2010}.

Optical measurements were performed using an external cavity diode laser (Toptica CTL 1500) and erbium-doped fiber amplifier (PriTel). In PL and PLE experiments, the signal was detected by a high-gain InGaAs photodiode (Femto OE-200); mechanical chopping (Thorlabs MC2000B) of both the excitation and detection arms was essential to minimize scattered excitation light at the detector. For the optical-ESR measurements the excitation light was delivered via a multimode optical fiber. The sample and the fiber were both fixed to a piece of sapphire with GE varnish to maintain consistent illumination. 

\begin{table*}[h!]
	\centering
	\begin{tabular}{|l|l|l|}
	\hline
	Implantation Energy / keV & Fluence (High Dose) / cm$^{-2}$ &  Fluence (Low Dose) / cm$^{-2}$ \\ \hline 
	10 & $3.8 \times 10^{10}$ & $3.8 \times 10^{12}$ \\ \hline
	25 & $5 \times 10^{10}$ & $5 \times 10^{12}$ \\ \hline
	50 & $7.5 \times 10^{10}$ & $7.5 \times 10^{12}$ \\ \hline
	100 & $1 \times 10^{11}$ & $1 \times 10^{13}$ \\ \hline
	150 & $1.3 \times 10^{11}$ & $1.3 \times 10^{13}$ \\ \hline
	250 & $1.3 \times 10^{11}$ & $1.3 \times 10^{13}$ \\ \hline
	350 & $3.8 \times 10^{11}$ & $3.8 \times 10^{13}$ \\ \toprule
	\textit{Total} & $9.0 \times 10^{11}$ & $9.0 \times 10^{13}$ \\ \hline
	\end{tabular}
	\caption{Implantation recipe for high and low dose sample}
	\label{tab:implantation}
\end{table*}

\section{Annealing conditions}

The samples were annealed in air in a tube furnace (Lindberg/Blue M).  The annealing temperature was reached by ramping the temperature at a rate of $\approx 5\degree $C/min, after which the temperature was kept constant during 2h, before cooling down to room temperature. This annealing process resulted in no change in the color and transparency of the samples, while annealing at the same temperatures in vacuum darkened the sample, consistent with literature reports of \ti{} reduction from the creation of oxygen vacancies~\cite{Khomenko1998}

\section{Implantation yield measurement}

We estimate the number of \er{} spins on the \ti{} site using quantitative ESR measurements. We compare the ESR response of \er{}:\tio\ with a control sample of \er:YSO (10 ppm, single crystal). By measuring the control and test samples simultaneously, we ensure that the temperature, cavity quality factor, and microwave power are the same. For all measurements, we also checked that the ESR spectra were taken in the linear regime with respect to microwave power and that the lines were not broadened by the magnetic field modulation. This allows us to quantitatively relate the ESR signal to the number of spins~\cite{Barr2008}.

\begin{figure}
    \centering
    \includegraphics[width=3.25in]{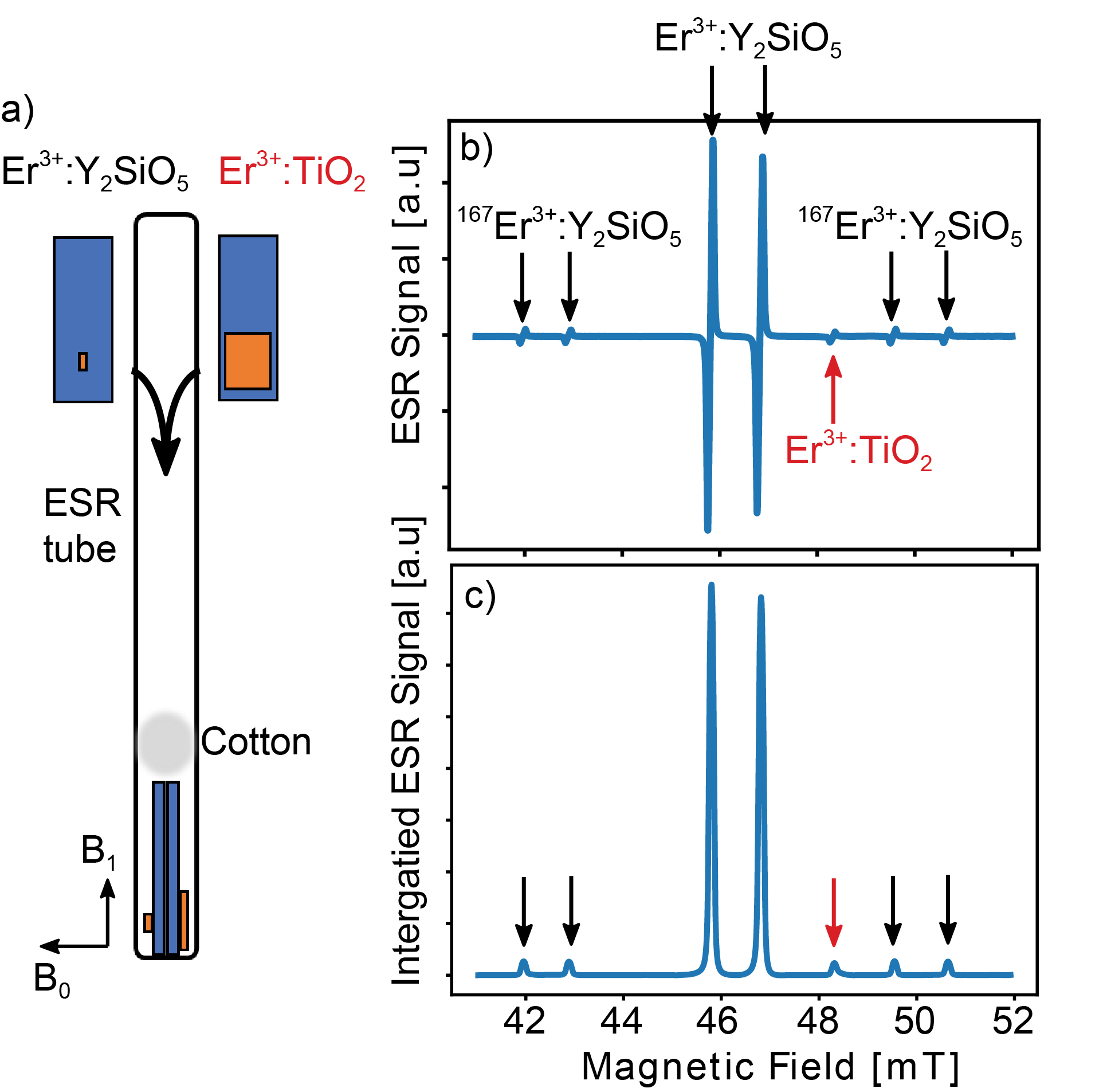}
    \caption{a) Sample arrangement in the ESR tube. b) ESR spectrum at 8 K with both the test and control samples in the cavity. Lines corresponding to \er:YSO and \er:\tio\ are labeled. The magnetic field is oriented along the $c$-axis of the \tio\ sample. c) Integrated spectrum corresponding to the ESR spectrum displayed in b).}
    \label{spin_counting_glyA2}
\end{figure}

The two samples are fixed to two different pieces of single-crystal quartz held in place by a piece of cotton (Figure \ref{spin_counting_glyA2}a)). We verified that the quartz and cotton pieces do not induce any perturbation signal at the magnetic fields used for the measurements. The temperature was maintained at 8 K, which ensures that $>$99.9\% of the \er\ population is in the lowest crystal field level, so thermal occupancy to the $Z_2$ level is ignored.

The control sample and the orientation of the samples in the spectrometer was the same for all measurements. The samples are oriented so that $\mathbf{B_0}$ points along the $c$-axis of the \tio\ sample and close to $D_1$ axis of the YSO sample. This orientation makes the effective $g$-factors of \er{} ions in \tio\ and in site 2 of YSO similar ($g_{TiO_2,||} \approx 14.3$ and $g_{YSO,||} \approx 15.1$). $\mathbf{B_0}$ is slightly misaligned ($<$5\degree\ in the $D_1b$ plane) from $D_1$ axis of \yso{} in order to split the two magnetically nonequivalent subsites in \yso{}~\cite{Sun2008}. Figure \ref{spin_counting_glyA2}b) shows an ESR spectrum where the signal from both samples is visible. The two strongest lines correspond to the $I = 0$ \er{} isotopes in site 2 in YSO. The other four lines labeled with black arrows correspond to the $\ket{-1/2,-1/2} \leftrightarrow \ket{1/2,-1/2}$ and $\ket{-1/2,1/2} \leftrightarrow \ket{1/2,1/2}$ ($\ket{m_S,m_I}$ states) transitions of \eri\ ions in the same site. The line labeled with a red arrow corresponds to the $I=0$ \er\ isotopes present in \tio. The oscillating magnetic field $\mathbf{B_1}$ in the cavity points along a direction close to [100] for the \tio\ sample and close to $D_2$ for the YSO, for which the effective $g$-factors are $g_{TiO_2,\perp} = 1.63\ (+0.4/-0)$ and $g_{YSO,\perp} = 1.86\ (+0.8 /-0.2)$ for the two samples, including the 5\degree\ misorientation.\\

{}

The mass of the control sample was measured to be $m = 1.2$ mg using a 0.1 mg precision balance. The density of YSO is $d = 4.44$ g.cm$^{-3}$. The total concentration of \er\ ions into the YSO piece is $c = 1.85 \times 10^{17}$ cm$^{-3}$. The number of \er\ ions with $I=0$ present in each subsite in YSO is therefore :
\begin{equation}
n_{YSO,I=0} = \frac{1}{2}\times\frac{1}{2}\times(1-\eta)\times c\times\frac{m}{d}
\end{equation}{}
with $\eta = 0.23$ the isotopic abundance of \eri. The two $\frac{1}{2}$ factors come from the fact that each site and each subsite is equally populated in YSO{}~\cite{Bottger2006}.

The double integral of the \er\ peak corresponding to the $I = 0$ isotopes of the measured sample (\tio) was compared to both of the \er\ $I = 0$ isotopes peaks of the control sample (YSO). An example of an integrated spectrum shown on Figure \ref{spin_counting_glyA2}c). The area of the $I = 0$ peaks of the YSO and \tio\ samples are determined and respectively called $A_{YSO,I=0}$ and $A_{TiO_2,I=0}$. 
The number of spins in the \tio\ samples can be calculated using the expression~\cite{Abragam1970}:
\begin{equation}
n_{TiO_2,I=0,measured} = \frac{A_{TiO_2,I=0}}{A_{YSO,I=0}}\times (\frac{g_{YSO,\perp}}{g_{TiO_2,\perp}})^2\times n_{YSO,I=0}
\end{equation}{}

The conversion efficiency is calculated as the ratio $\frac{n_{TiO_2,I=0,measured}}{n_{TiO_2,I=0,implanted}}$.\\

The number of implanted spins $n_{TiO_2,I=0,implanted}$ is estimated using the implantation dose multiplied by the surface of the sample and by the isotopic abundance $(1-\eta)$. The conversion efficiencies for the different samples are displayed in the table 2 in the main text.

The error sources in measuring conversion efficiency are listed in table \ref{error_estimations}. The same reference sample was used in all cases, so uncertainty in the mass, \er\ concentration and $g_{TiO_2,\perp}$ can be neglected when considering the relative uncertainty between samples.

\begin{table*}[]
    \centering
    \begin{tabular}{|c|c|c|c|}
    \hline
    Parameter & Value & Total error & Uncorrelated error \\
    \hline \hline
    Mass of \yso & 1.2 mg & 0.08 & 0\\
    \hline
    \er concentration in \yso & 10 ppm & 0 & 0\\
    \hline
    g$_{TiO_2,\perp}$ & 1.63 & 0.20 & 0.20\\
    \hline
    g$_{YSO,\perp}$ & 1.86 & 0.20 & 0 \\
    \hline
    A$_{YSO,I=0}$ & Variable & 0.10 & 0.10 \\
    \hline
    A$_{TiO_2,I=0}$ & Variable & 0.10 & 0.10 \\
    \hline
    \er\ implantation dose & Variable & 0.10 & 0 \\
    \hline
    Sample area & Variable & 0.10  & 0.10\\
    \hline
    \hline
    Conversion efficiency & Variable & 0.39 & 0.28  \\
    \hline
    \end{tabular}
    \caption{Error estimations for the different parameters taken into account for conversion efficiency. Here, total error denotes the fractional uncertainty in each parameter affecting the absolute accuracy of all the measurements, while uncorrelated error denotes the fractional uncertainty contributing to relative errors between different samples.}
    \label{error_estimations}
\end{table*}{}

\bibliography{library_er_tio2}